\documentclass[twocolumn]{aastex631}

\usepackage{graphicx}
\usepackage[none]{hyphenat}
\usepackage{amsmath}
\usepackage{booktabs}
\usepackage{multirow}
\usepackage{txfonts}
\usepackage{enumitem}

\usepackage{algorithmic}
\usepackage{bm}
\usepackage{mathrsfs}
\usepackage{booktabs}  
\usepackage{float}
\usepackage{subfigure}
\DeclareRobustCommand{\VAN}[3]{#2}
\let\VANthebibliography\thebibliography
\def\thebibliography{\DeclareRobustCommand{\VAN}[3]{##3}\VANthebibliography}
\usepackage{color}

\shorttitle{Cascade Processes of Strong and Weak MHD Turbulence}
\shortauthors{Gao, Zhang \& Cho}

\begin{document}
\title{Cascade processes of strong and weak relativistic magnetohydrodynamic turbulence}

\author{Na-Na Gao}
\affiliation{Department of Physics, Xiangtan University, Xiangtan, Hunan 411105, China}

\author{Jian-Fu Zhang}
\affiliation{Department of Physics, Xiangtan University, Xiangtan, Hunan 411105, China}
\affiliation{Key Laboratory of Stars and Interstellar Medium, Xiangtan University, Xiangtan 411105, China}
\email{jfzhang@xtu.edu.cn}

\author{Jungyeon Cho}
\affiliation{Department of Astronomy and Space Science, Chungnam National University, Daejeon, Republic of Korea}
\affiliation{School of Physics Science and Technology, Kunming University, Kunming 650214, China}
\email{jcho@cnu.ac.kr}

\begin{abstract}
On the framework of relativistic force-free magnetohydrodynamic (MHD) turbulence, we explore the fundamental properties of strong and weak turbulent cascades using high-resolution numerical simulations in the presence of a uniform background magnetic field. We find that (1) power spectra and scale-dependent anisotropies both for the strong and weak turbulence resemble those observed in the non-relativistic MHD turbulence; (2) intermittency of magnetic fields in strong turbulence is stronger than that in the weak one; (3) generated Alfv\'en modes show similar energy spectra and scale-dependent anisotropies to those of non-relativistic case; (4) generated fast modes present a power spectrum similar to that of Alfv\'en modes, with a strong (for strong turbulence) or weak (for weak turbulence) scale-dependent anisotropy, which are significantly different from non-relativistic turbulence; and (5) applications of our numerical results to neutron star magnetospheres show that the strong (or moderately weak) turbulent cascade can explain the X-ray radiation of the Vela pulsar. Our study is of great significance for understanding energy transfer, magnetic field evolution, and particle acceleration mechanisms in extreme astrophysical environments.
\end{abstract}


\section{Introduction}\label{sec:intro}
Magnetohydrodynamic (MHD) Turbulence plays an important role in many key astrophysical processes. The standard nonrelativistic MHD (NRMHD) turbulence, relevant to diffusive interstellar media, turbulent reconnection, and the transport of cosmic rays, has been extensively studied from both theoretical (\citealt{Goldreich1995}; hereafter GS95) and numerical perspectives (\citealt{Cho2000, Maron2001, Cho2002}). Under the condition of a critical balance assumption (the Alfv\'enic wave propagation timescales equal to the eddy turnover times), GS95 predicted that the strong Alfv\'enic turbulence has a Kolmogorov energy spectrum of $E(k_\perp) \propto k_\perp^{-5/3}$ (\citealt{Kolmogorov1941}), and a scale-dependent anisotropy of $k_{\parallel} \propto k_{\perp}^{2/3}$, where $k_{\parallel}$ and $k_{\perp}$ are parallel and perpendicular wavenumbers with regard to the local mean magnetic field direction, respectively. These theoretical predictions were numerically verified by \cite{Cho2000}. Subsequently, compressible MHD turbulence was further decomposed numerically by \cite{Cho2002, Cho2003} into three plasma modes, i.e., Alfv\'en, fast, and slow modes. For the weak Aflv\'enic turbulence (here, the so-called ``weak'' means that nonlinear energy cascade timescale is longer than the wave period), it was demonstrated that the turbulence spectra adhere to a scaling law of $E(k_{\perp}) \propto k_{\perp}^{-2}$ (\citealt{Galtier2000, Kuznetsov2001, Perez2008}). However, there are still many open issues (see \citealt{Schekochihin2022} for more details), such as the imbalance problem between the velocity field and the magnetic field (i.e., the residual energy problem; \citealt{Boldyrev2012, Chandran2015}), the characteristics of MHD turbulence at the sub-viscous scale (\citealt{Cho2003b}), physical nature of intermittent dimensions (\citealt{Meyrand2015}), the weak-to-strong transition of MHD turbulence (\citealt{Meyrand2016, Zhao2024}), and the anisotropy of weak turbulence (e.g., \citealt{Galtier2000, Beresnyak2019}).

The relativistic MHD (RMHD) turbulence is relevant to relativistic astrophysical sources such as the magnetospheres of pulsars, gamma-ray bursts, black hole accretion disks, and jets. At present, there is still no theoretical prediction of the fully RMHD turbulence. Numerically, simulations of the RMHD turbulence (\citealt{Zrake2012, Takamoto2017}) showed that the power spectrum is compatible with a Kolmogorov spectrum: $E(k) \propto k^{-5/3}$. In particular, they observed the anisotropic scaling of $k_{\parallel} \propto k_{\perp}^{0.84}$ and $k_{\perp}^{0.7}$ in weakly (\citealt{Zrake2012}) and strongly (Poynting flux dominated; \citealt{Takamoto2017}) magnetized RMHD, respectively, the former of which is steeper than the relation of $k_{\parallel} \propto k_{\perp}^{2/3}$ in NRMHD turbulence (GS95). Additionally, \cite{Zrake2012} found a similar intermittency behavior with the nonrelativistic case predicted by \cite{She1994} in longitudinal velocity fluctuation and a stronger intermittency behavior in the transverse component. Unlike the NRMHD turbulence, RMHD turbulence shows a strong coupling between the fast and Alfv\'en modes (\citealt{Takamoto2016}), resulting in the spectrum of fast mode being the same as that of Alfv\'en mode. We noted that the above results are from the velocity field in the case of strong turbulence. Currently, we do not know whether the characteristics of the turbulent magnetic field are similar to those of velocity fields, and the specific properties of weak turbulence remain unknown. Besides, some key differences between RMHD and NRMHD turbulence lie in the former being constrained by high magnetization, the speed limit of light, and energy-momentum coupling, resulting in suppression of slow waves, limited generation of fast waves, magnetic energy dominance, enhanced resistive dissipation, and more challenging numerical implementation. The latter exhibits greater diversity in the kinetic and magnetic energy balance, the contributions of compressible modes, and dissipation mechanisms.

In a highly magnetized and Poynting flux-dominated environment, the relativistic force-free MHD is a limit applicable to relativistic astrophysical sources,  where the energy density of matter is negligible compared to electromagnetic energy density. Exploring the properties of force-free MHD turbulence is crucial for understanding various astrophysical sources and processes such as pulsar and black hole magnetospheres (\citealt{Goldreich1969, Blandford1977, Duncan1992, Komissarov2002}), gamma-ray bursts (\citealt{Thompson1994, Lyutikov2003}), turbulent magnetic reconnection (\citealt{Ripperda2021, Liang2023, Zhang2023, Liang2025}), particle transport and acceleration (\citealt{Zhang2021, Gao2024, Gao2025, Xiao2025}). In the context of extreme relativistic MHD in the force-free approximation, \citet[hereafter TB98]{Thompson1998} theoretically predicted that the strong force-free Alfv\'enic turbulence exhibits a Kolmogorov spectrum and a Goldreich-Sridhar type anisotropy, which closely resembles its nonrelativistic counterpart (GS95), and the weak force-free Alfv\'enic turbulence displays an energy spectrum characterized by $E(k) \propto k^{-2}$, similarly akin to the nonrelativistic case (\citealt{Galtier2000, Kuznetsov2001}). 

For the strong force-free Alfv\'enic turbulence, \citet[hereafter Cho05]{Cho2005} was the first to numerically confirm the theoretical predictions of TB98, albeit with low numerical resolution (see also \citealt{Cho2014} for strong imbalanced force-free Alfv\'enic turbulence). As far as we know, there is still no direct numerical testing of TB98's predictions in the case of weak force-free Alfv\'enic turbulence. Slightly similar to this context, recent simulations by \cite{Ripperda2021} show that nonlinear interactions between two overlapping, perpendicularly polarized Alfv\'en waves result in significant dissipative structures of current sheets that have an energy spectral characteristic of weak turbulence, namely $E(k_{\perp}) \propto k_{\perp}^{-2}$.

Interestingly, we note that \cite{Li2019} investigated Alfv\'en wave dissipation in force-free electrodynamic simulations of relativistic magnetospheres of magnetars, characterizing the driving mode by the initial wave amplitude $b/B_0$, where $B_0$ and $b$ are the mean and disturbance magnetic field strength, respectively. Although they claimed a spectrum of $E(k_{\perp}) \propto k_{\perp}^{-2}$ (see Fig. 4 of their paper), it looks like a steeper spectral index than $-2$ to the naked eye. Substituting their initial wave amplitudes $b/B_0 = [0.5, 3]$ and driving wavenumber $k_{\perp}=k_{\parallel}$ into the critically balanced condition $\chi \equiv b k_{\perp}/B_0 k_{\parallel}$ (see the following Eq. (\ref{eq.criticalbalance})) used in our work, we obtain $\chi = [0.5, 3]$, corresponding to the transition regime of weak-strong turbulence. Theoretically, the expected spectral exponents, however, should be $-2$ and $-5/3$ respectively in the case of $\chi<1$ and $\chi\geq1$. We speculate that the steeper spectral index shown in \cite{Li2019} might correspond to not fully developed turbulence. 

Given the above mentioned open issues related to RMHD turbulence, this work is to elucidate the fundamental properties of strong and weak force-free MHD turbulence via high-resolution simulations, focusing on the energy spectrum, anisotropy, and intermittency of RMHD turbulence, as well as the characteristics of plasma modes. Applying our numerical results to the magnetospheres of neutron stars, we also discuss the properties of RMHD turbulence cascade to understand the energy transfer and acceleration mechanisms in neutron stars. This paper is structured as follows. We describe the numerical methods and outline our initial simulation setup in Section \ref{sec:numre.methods}. Numerical results are presented in Section \ref{sec:numre.results}. We discuss applications of our results in Section \ref{sec:application-NS}, using the Vela pulsar as an example. Sections \ref{sec:discu} and \ref{sec:summa} provide the discussion and summary, respectively.

\section{Numerical Methods}\label{sec:numre.methods}
Following Cho05, we numerically solve the system of RMHD equations (see Eqs. (1) to (6) of Cho05), using a MUSCL-type scheme with HLL fluxes (\citealt{Harten1983}) and monotonized central limiter (\citealt{Kurganov2001}). The second-order accurate flux-interpolated constrained transport scheme is adopted to maintain the condition of $\nabla \cdot {\bm B}=0$ (\citealt{Toth2000}). Meanwhile, we keep the force-free MHD conditions of 
\begin{equation}
{{\bm E} \cdot {\bm B}} = 0,\ \ \  {\rm and} \ \ \  B^2 - E^2 > 0. \label{eq.condition}
\end{equation}
met at every time step, where $\bm B$ and $\bm E$ are the magnetic and electric field, respectively. Our simulations take place on the periodic cube of size $2\pi$, with a high resolution of $1024^3$. The magnetic field is composed of a uniform background field and a fluctuating field, $\bm {B} = {\bm B_0} + {\bm b}$ (setting $B_0=1$ in this work), while the electric field consists only of a fluctuating term. We fix the initial pressure $p_{0}=0.1$ and density $\rho_{0}=1.0$, leading to the initial plasma parameter $\beta_0 = 2p_0/B_{0}^{2} = 0.2$.

The force-free MHD turbulence admits two normal waves, Alfv\'en and fast waves (see also TB98; \citealt{Komissarov2002}). By analytically exploring nonlinear interactions between these plasma waves, TB98 predicted that two counter-propagating Alfv\'en waves can generate either Alfv\'en or fast waves through wave-wave interactions, $\mathcal{A}^{+} + \mathcal{A}^{-} \rightarrow \mathcal{A}^{+} + \mathcal{A}^{-}$ and $\mathcal{A}^{+} + \mathcal{A}^{-} \rightarrow \mathcal{F}$. In addition, an incoming Alfv\'en wave and an incoming fast wave can generate either an Alfv\'en or a fast outgoing wave ($\mathcal{A} + \mathcal{F} \rightarrow \mathcal{A} / \mathcal{F}$). If two incident fast waves interact, a fast wave will be generated ($\mathcal{F} + \mathcal{F} \rightarrow \mathcal{F}$), without involving Alfv\'en waves. In this paper, we consider the Alfv\'enic turbulence interactions between $\mathcal{A}^{+}$ and $\mathcal{A}^{-}$. At the beginning of the simulation, we excite Alfv\'en modes in the wavenumber range of (see also Cho05)
\begin{equation}
1 \le k_{\parallel} \ge 2,\ \ \  {\rm and} \ \ \  4 \le k_{\perp} \ge 6. \label{eq.drive-k} 
\end{equation}

To explore the properties of strong and weak force-free MHD turbulence, we conduct three group high-resolution simulations: one strong turbulence of $\chi \sim 1.0$, and two weak cases of $\chi \sim 0.5$ and $\chi \sim 0.25$, through regulating the amplitude of magnetic field disturbance $b/B_0$ in the critically balanced condition
\begin{equation}
\chi \equiv \frac{b k_{\perp}}{B_0 k_{\parallel}}. \label{eq.criticalbalance}
\end{equation}
When $\chi \sim 1.0$, the resulting turbulence is strong (see GS95 for more discussion about non-relativistic Alfv\'enic turbulence); when $\chi < 1.0$, the turbulence is weak (see TB98 for more discussions about weak force-free MHD turbulence; see also \cite{Galtier2000} for the non-relativistic case). 

\begin{figure}
\centering
\includegraphics[width=0.9\columnwidth,height=0.18\textheight]{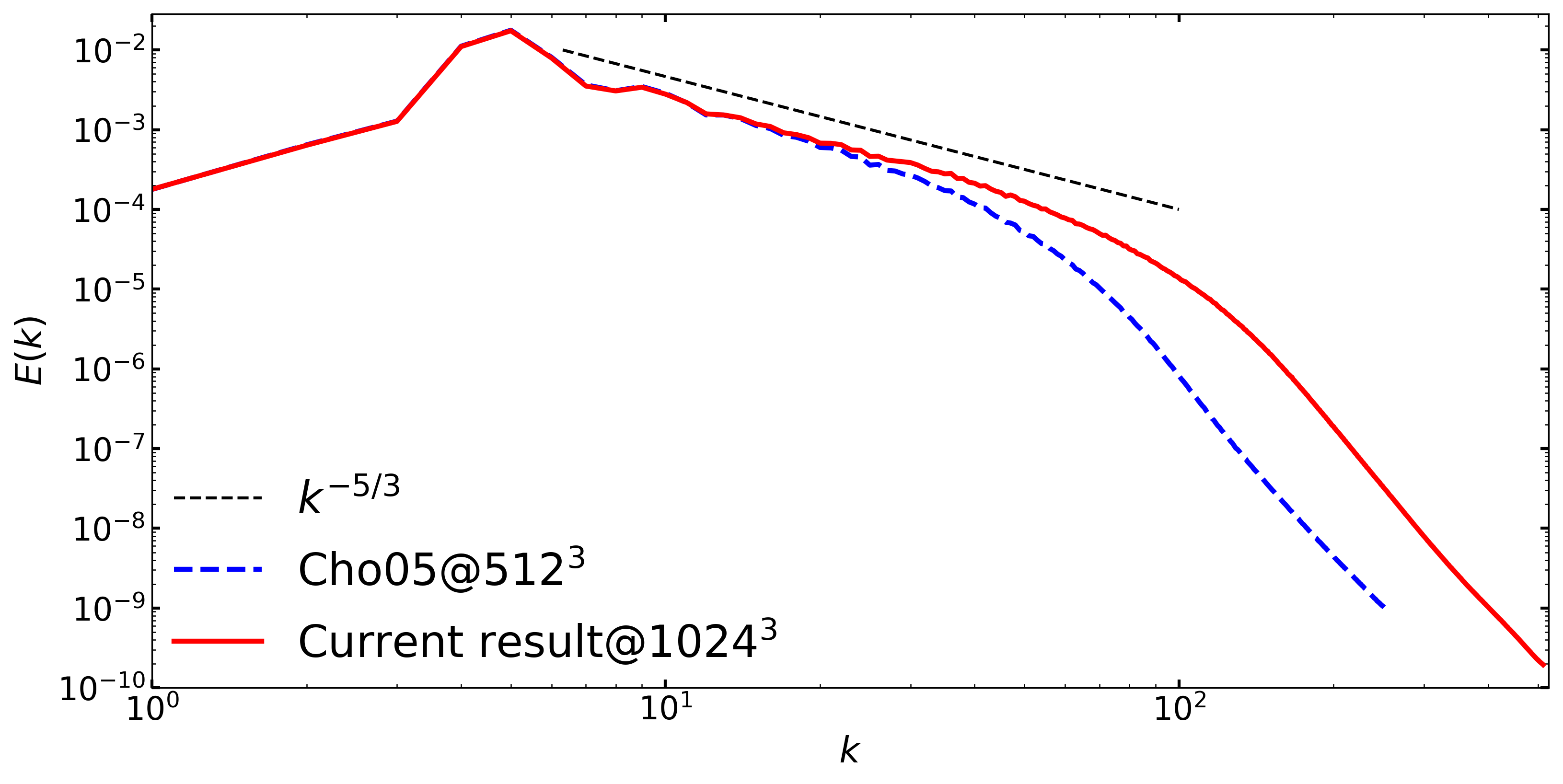}
\caption{Comparison of magnetic field power spectrum with different numerical resolutions at $t \sim 5.86$ for $\chi \sim 1.0$.
}
\label{fig_ps_resolu}
\end{figure}

\section{Numerical Results}\label{sec:numre.results}
\subsection{Comparison with Earlier Results}\label{sec:numre.compaison}
Figure \ref{fig_ps_resolu} compares the magnetic field power spectrum for numerical resolutions of $1024^3$ to that of $512^3$ (same as the result of Cho05) at $t \sim 5.86$, where turbulence has fully developed (see Fig. \ref{fig_pst} for more details). We can clearly observe that the spectra are compatible with a Kolmogorov spectrum of $E(k) \propto k^{-5/3}$. Compared with the $512^3$ simulation (consistent with that of Cho05), we find that the $1024^3$ simulation presents an extended inertial (power-law) range. When simulating a weak force-free MHD turbulence (the focus of the current work), we realize that it is difficult to measure an inertial range with a low-resolution simulation. As seen in the following Figs. \ref{fig_pst} and \ref{fig_ps}, weak turbulence exhibits a larger dissipation scale than strong turbulence, that is, the former has a narrower inertial range than the latter\footnote{Our simulations used the ideal force-free MHD equations, which contain no explicit physical magnetic diffusivity. Therefore, for a fixed numerical setup, the effective magnetic diffusivity $\eta$ is a constant, which is identical for all our runs with different $\chi$. Theoretical scalings such as $\ell_{\rm d} \sim \eta^{3/4} \epsilon^{-1/4}$ and $\ell_{\rm d} \sim (\eta / |\dot{E}|)^{1/2}$ predict a larger $\ell_{\rm d}$ for weaker turbulence due to its lower cascade efficiency (smaller $\epsilon$) and longer decay timescale (smaller $|\dot{E}|$). This is confirmed by our direct measurement from the magnetic energy spectra in Figure 4(a), where the spectral cutoff (hence $\ell_{\rm d}$) increases as $\chi$ decreases.}. In particular, a high-resolution simulation is necessary to explore the spatial structure (which can be revealed by anisotropy and intermittency) of relativistic force-free MHD turbulence.

\subsection{Evolution Processes of the force-free MHD Turbulence}\label{sec:numre.evolu}
\begin{figure}
\centering
\includegraphics[width=0.9\columnwidth,height=0.45\textheight]{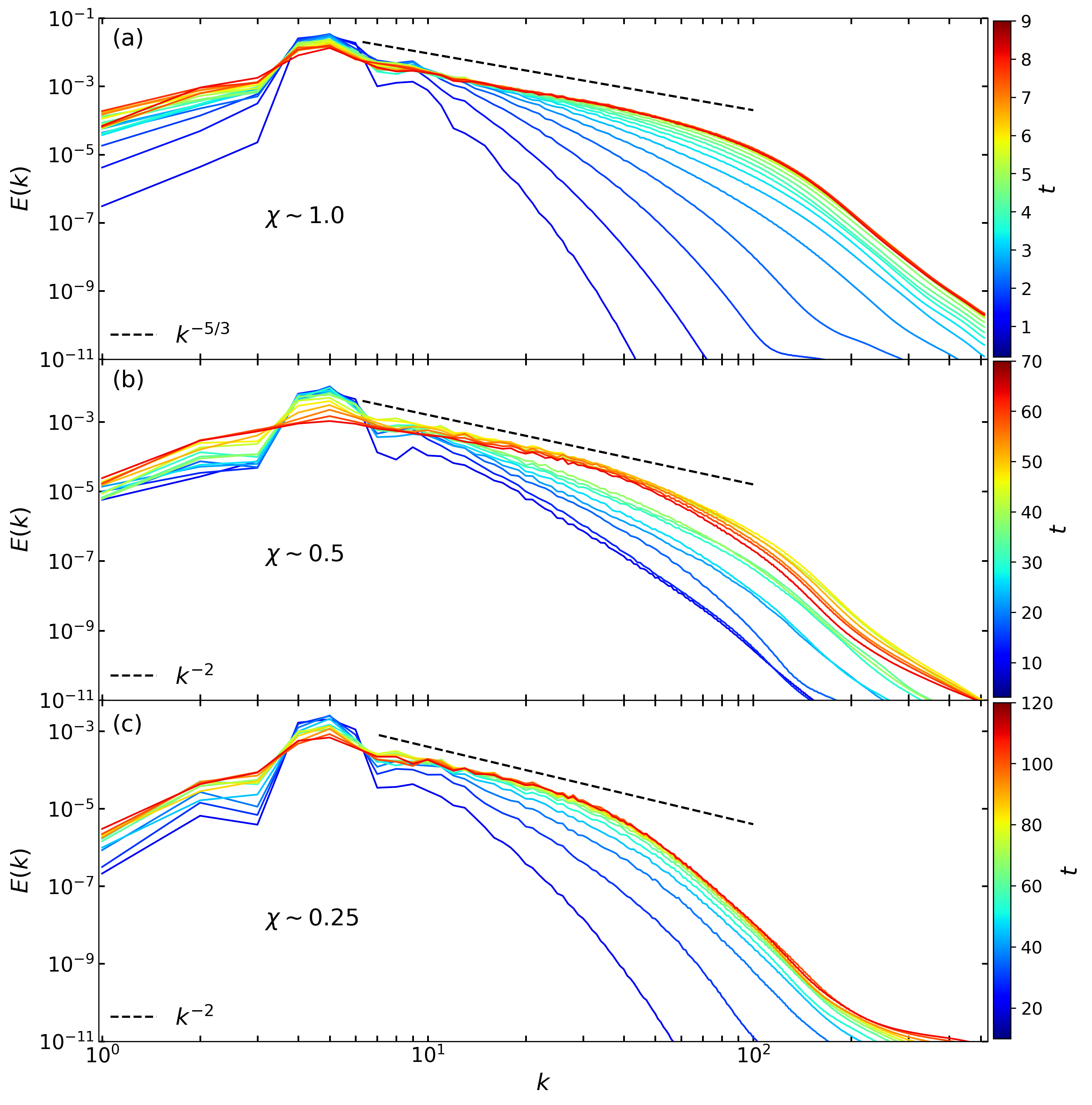}
\caption{The time-dependent evolution of magnetic field power spectrum for the cases of $\chi \sim 1.0$ (panel (a)), 0.5 (panel (b)) and 0.25 (panel (c)).
}
\label{fig_pst}
\end{figure}

We show the time-dependent magnetic field power spectra for $\chi \sim 1.0$, 0.5, and 0.25 in panels (a), (b), and (c) of Fig. \ref{fig_pst}, respectively. As is seen, the magnetic energy spectra over time have a similar evolution trend in these three cases: (1) at $t = 0$, large-scale (small $k$) Fourier modes are excited only (not shown here), due to the fact that we initially drive the Alfv\'en waves at large scale (see Eq. (\ref{eq.drive-k})); (2) as time evolves, energy cascades down to small-scale (large $k$) modes, in the form of a steeper power law; (3) at the later stage of evolution, the energy spectrum remains a constant slope: $-5/3$ for strong turbulence (see panel (a) for $\chi \sim 1.0$), and $-2$ for weak turbulence (see panels (b) and (c) for $\chi < 1.0$). From Fig. \ref{fig_pst}, it is evident that turbulence fully develops at approximately $t \sim 5.0$, 45, and 100 for $\chi \sim 1.0$, 0.5, and 0.25, respectively, which indicates that strong turbulence develops significantly more rapidly than weak turbulence. Additionally, as the value of $\chi$ decreases, the time required for weak turbulence to be fully developed increases. This trend of faster development for strong turbulence is consistent with that observed in NRMHD turbulence (e.g., \citealt{Galtier2000, Cho2004}). Subsequent analysis of all turbulence properties will be conducted based on times $t \sim 9.0$, 50, and 120 for $\chi \sim 1.0$, 0.5, and 0.25, respectively.

\subsection{Properties of fully developed turbulence}\label{sec:numre.properties}
\subsubsection{Probability Distribution Function}\label{sec:numre.prop}

\begin{figure}
\centering
\includegraphics[width=0.9\columnwidth,height=0.2\textheight]{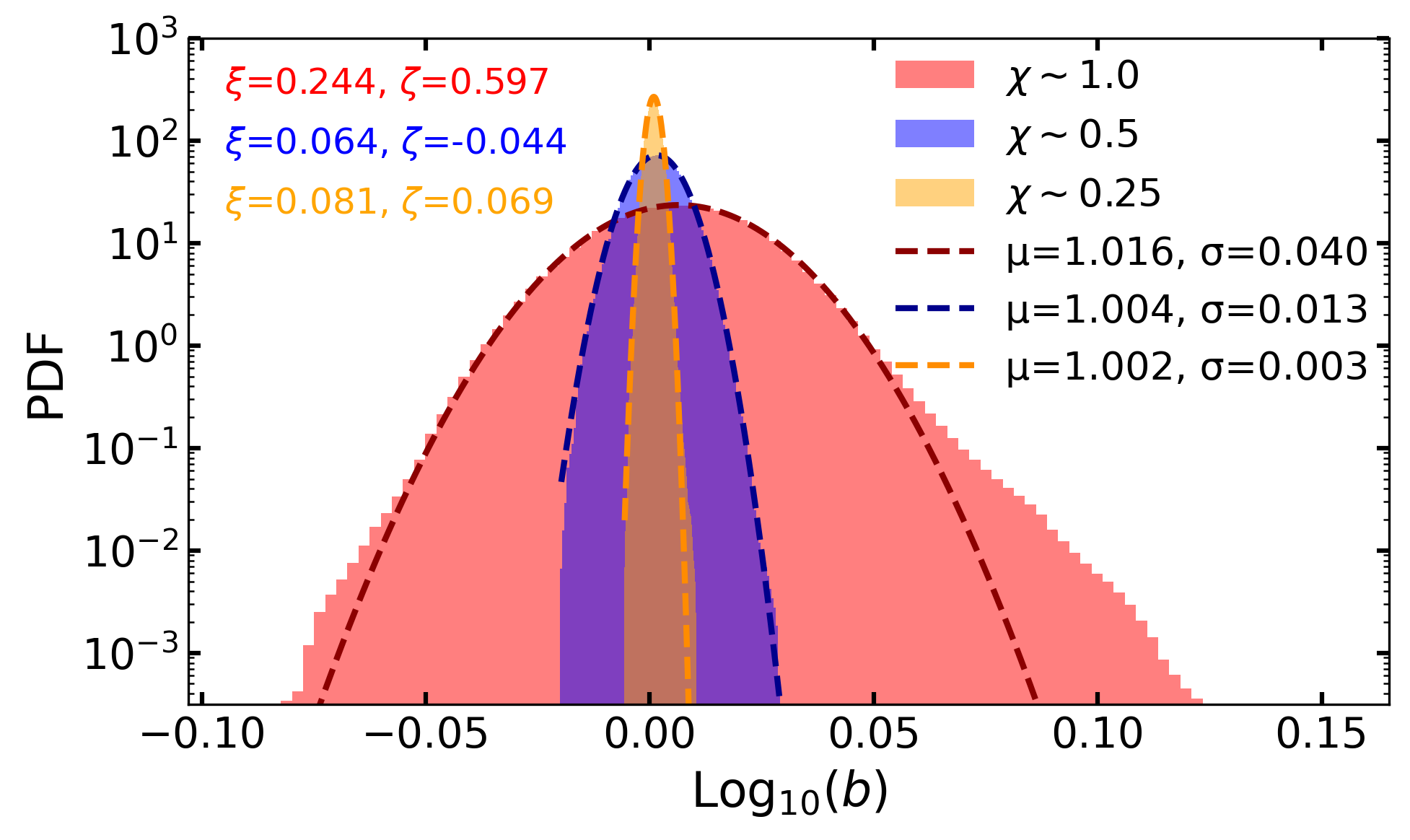}
\caption{The probability distribution function of the magnetic field for $\chi \sim 1.0$ (red), 0.5 (blue), and 0.25 (orange). Gaussian fitting results with the mean value $\mu$ and standard deviation $\sigma$ are plotted as dashed lines in the corresponding colors. $\xi$ and $\zeta$ represent higher-order statistical measurements -- skewness and kurtosis, respectively.
}
\label{fig_pdf}
\end{figure}

Based on fully developed turbulence, we analyze the probability distribution function (PDF) of the magnetic field for the three cases, as shown in Fig. \ref{fig_pdf}, providing the values of the statistical moments of each order. Although the analysis of the low-order statistical moments ($\mu$ and $\sigma$) can hardly distinguish the statistical characteristics between the strong and weak relativistic force-free MHD turbulence, the high-order statistical moments (skewness $\xi$ and kurtosis $\zeta$) have significant differences for $\chi \sim 1.0$ and $\chi < 1.0$. The values of $\xi$ and $\zeta$ in weak turbulence are much smaller than those in strong turbulence, which indicates that the probability distributions for $\chi \sim 1.0$ have more elongated non-Gaussian tails (a clear signature of intermittency) and flatter distributions than those for $\chi < 1.0$. As a result, our results reveal that relativistic weak turbulence is obviously different from relativistic strong turbulence. This trend of stronger intermittency in strong turbulence is consistent with findings in NRMHD turbulence (e.g., \citealt{Muller2000}).

\subsubsection{Spectra}\label{sec:numre.spec}

\begin{figure}
\centering
\includegraphics[width=0.9\columnwidth,height=0.5\textheight]{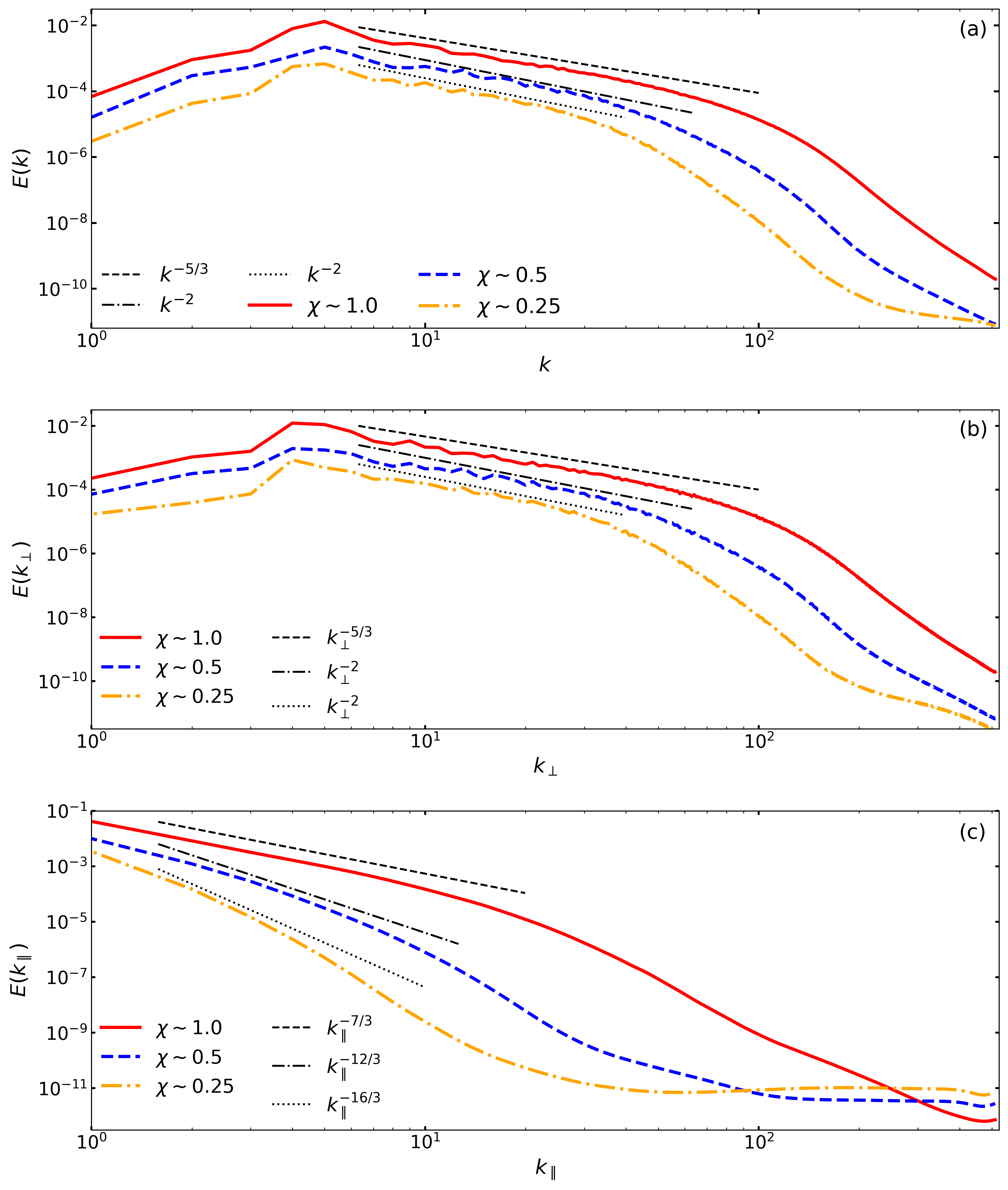}
\caption{The overall (panel (a)), perpendicular (panel (b)) and parallel (panel (c))  power spectra of the magnetic field.
}
\label{fig_ps}
\end{figure}

Figure \ref{fig_ps}(a) shows the power spectra of the magnetic at $\chi \sim 1.0$, 0.5, and 0.25 for fully developed turbulence. We can observe that: for strong turbulence ($\chi \sim 1.0$), the power spectrum presents a power-law relationship of $E(k) \propto k^{-5/3}$, in agreement with the earlier theoretical predictions for the non-relativistic (GS95) and relativistic (TB98) Alfv\'enic turbulence; for weak turbulence ($\chi \sim 0.5$ and 0.25), the power spectra show a power-law relationship of $E(k) \propto k^{-2}$, consistent with the theoretical results for the non-relativistic (\citealt{Galtier2000}) and relativistic (TB98) cases. 

To investigate the energy cascade properties in the relativistic strong and weak force-free MHD turbulence, we show the parallel (panel (b)) and perpendicular (panel (c)) magnetic field power spectra with respect to the local magnetic field in Fig. \ref{fig_ps}. We see that the perpendicular power spectra present the same power law as the overall spectra (see Fig. \ref{fig_ps}(a)), with $E_{b}(k_{\perp}) \propto k_{\perp}^{-5/3}$ and $k_{\perp}^{-2}$ in weak and strong turbulence, respectively. Differently, the parallel power spectra $E_{b}(k_{\parallel})$ show steeper power laws than $E_{b}(k_{\perp})$, with $E_{b}(k_{\parallel}) \propto k_{\parallel}^{-7/3}$, $k_{\parallel}^{-4}$ and $k_{\parallel}^{-16/3}$ for $\chi \sim 1.0$, 0.5 and 0.25, respectively. Therefore, the perpendicular cascade dominates the energy transfer of wave-wave interaction. In particular, for $\chi <1$, $E_{b}(k_{\parallel})$ remains very small in a wide wavenumber range, suggesting the negligible energy cascade in the direction parallel to the local magnetic field. Note that the amplitude of the perpendicular power spectra $E_{b}(k_{\perp})$ peak at $k_{\perp} \sim 4$, and the parallel ones $E_{b}(k_{\parallel})$ maximize at $k_{\parallel} \sim 1$, which are associated with the wavenumber range that we set Alfv\'enic wave initial driving (see Eq. (\ref{eq.drive-k})).

\subsubsection{Anisotropy}\label{sec:numre.prop.anis}
\begin{figure}
\centering
\includegraphics[width=0.9\columnwidth,height=0.25\textheight]{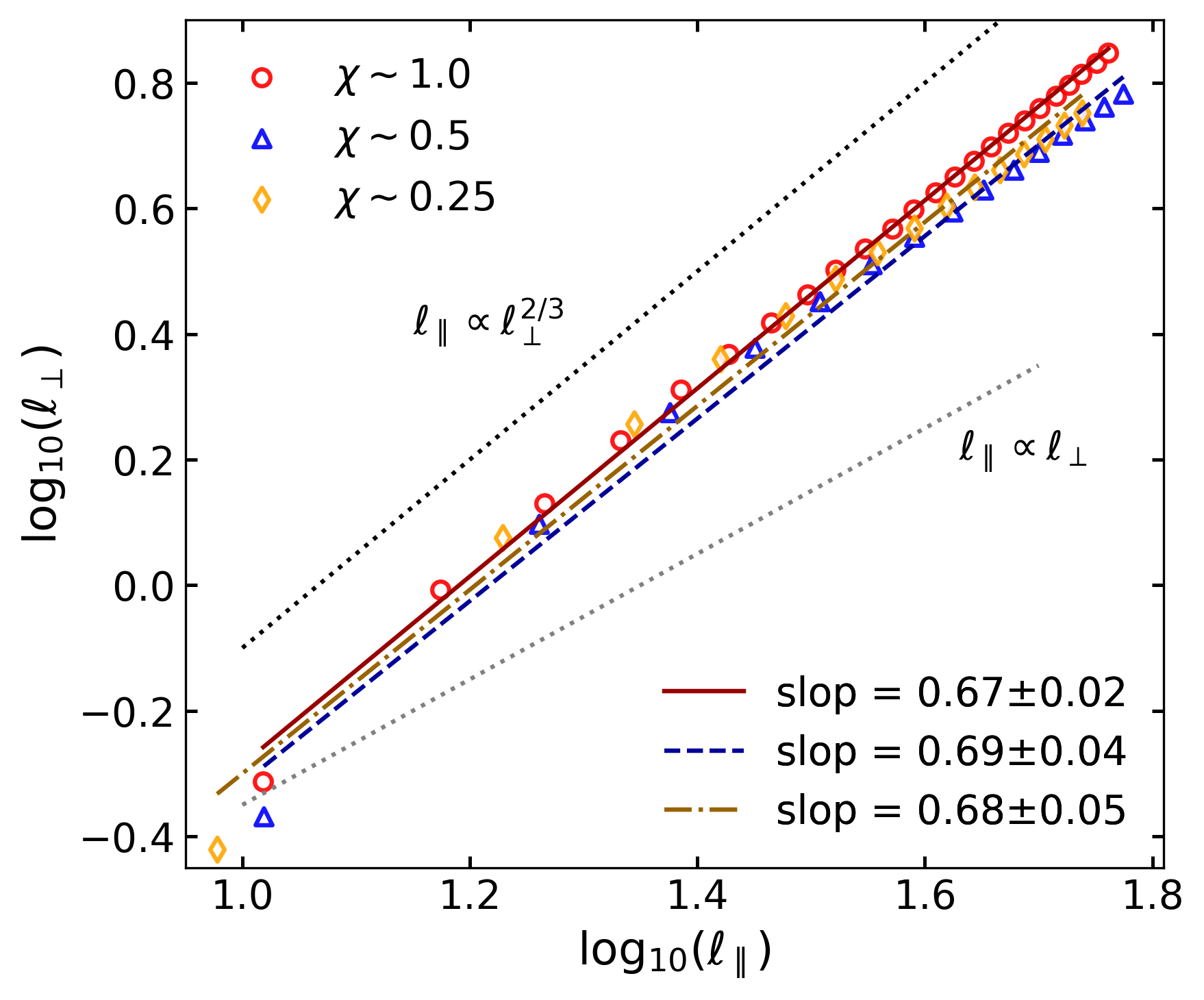}
\caption{Scale-dependent anisotropy of the magnetic field. The coordinates $\ell_{\perp}$ and $\ell_{\parallel}$ represent the perpendicular and parallel scales of the eddies along the local magnetic field, respectively. The colored lines are the corresponding linear fitting.
}
\label{fig_anisotropy}
\end{figure}

To investigate the anisotropy and intermittency (see Section \ref{sec:numre.prop.intermi}) of the relativistic force-free MHD turbulence, we here define the multi-order structure function as 
\begin{equation}
{\rm SF}_{p}(\bm R) = \langle {\mid F({\bm X} + {\bm R}) - F({\bm X}) \mid}^p \rangle, \label{eq.sf}
\end{equation}
for any fluctuation quantity $F$, where $\langle ... \rangle$ represents a spatial average of the system over the three-dimensional position vector $\bm X$. ${\bm R} = R_{\parallel} \hat{\bm R_{\parallel}} + R_{\perp} \hat{\bm R_{\perp}}$ is the spatial separation, where $\hat{\bm R_{\parallel}}$ and $\hat{\bm R_{\perp}}$ are unit vectors parallel and perpendicular to the local mean magnetic field (see \citealt{Cho2002} and \citealt{Cho2000} for detailed discussions), respectively. 

For the strong and weak turbulence, we first calculate the second-order structure function of the magnetic field and velocity (setting $p=2$ in Eq. (\ref{eq.sf})). By analyzing the relation between perpendicular sizes of eddies ($\sim 1/k_{\perp}$) and the parallel ones ($\sim 1/k_{\parallel}$), we then obtain a scale-dependent anisotropic scaling of $\ell_{\parallel} \propto \ell_{\perp}^{2/3}$ on small scales, as shown in Fig. \ref{fig_anisotropy}. This numerical finding indicates that: 1) there is a similar anisotropic relationship between relativistic strong and weak turbulence; 2) relativistic turbulence has a similar anisotropic scaling with non-relativistic turbulence (GS95; \citealt{Cho2002}).

\subsubsection{Intermittency}\label{sec:numre.prop.intermi}
\begin{figure*}
\centering
\includegraphics[width=1.9\columnwidth,height=0.25\textheight]{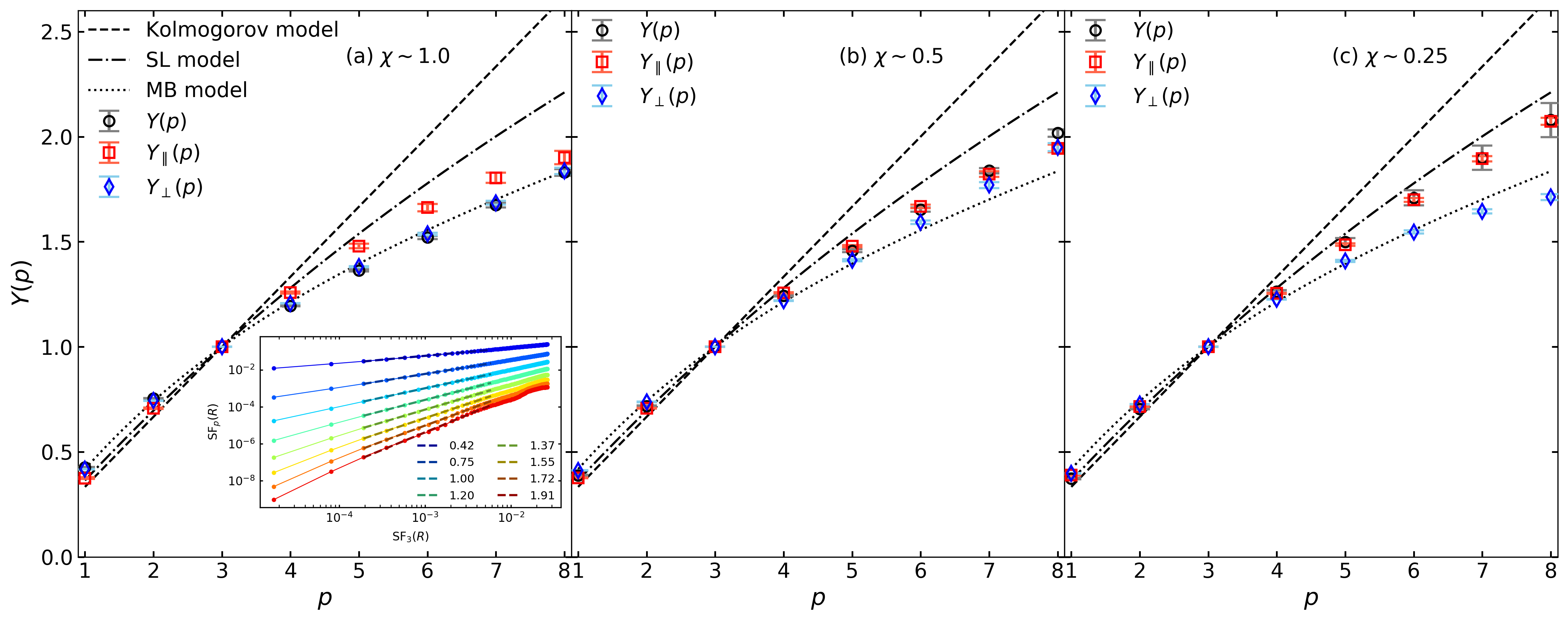}
\caption{Intermittency of magnetic fields for $\chi \sim 1.0$ (panel (a)), 0.5 (panel (b)) and 0.25 (panel (c)), shown as the scaling exponent $Y$ vs. the order of structure function $p$, including the total $Y(p)$, parallel $Y_{\parallel}(p)$ and perpendicular $Y_{\perp}(p)$ components. Our numerical results, with error bars estimated from the standard deviation, compare with the theoretical predictions of the Kolmogorov, SL, and MB models. The inset in panel (a) shows the multi-order structure function ${\rm SF}_{p}(\bm R)$ (from $p=1$ to $8$) vs. ${\rm SF}_{3}(\bm R)$ under the extended self-similarity hypothesis, and the colored dashed lines show the corresponding linear fitting. 
}
\label{fig_intermittency}
\end{figure*}

Intermittency is a manifestation of the deviation of the turbulent energy cascade from a uniform distribution, characterized by 
\begin{equation}
{\rm SF}_{p}(\bm R) = {\bm R}^{Y (p)}  \label{eq.sf-R}
\end{equation}
the scaling exponent of the multi-order structure function, where $Y(p)$ is the absolute scaling exponent related to the order of the structure function $p$ (normalized by the scaling exponent of the third-order structure function). Considering the incompressible hydrodynamic turbulence is self-similar within the inertial range, the absolute scaling exponent and the order of structure function exhibit a linear relationship of $Y(p) = p/3$ (\citealt{Kolmogorov1941}), called the Kolmogorov model. When the relation between $Y(p)$ and $p$ deviates from this Kolmogorov model, it indicates the occurrence of the intermittency phenomenon. As well known, \cite{She1994} and \cite{Muller2000} proposed classical nonlinear relations of $Y(p) = \frac{p}{9} + 2[1-(2/3)^{p/3}]$ to characterize the 1D vortex filament for incompressible non-relativistic hydroturbulence (referred to as SL model) and $Y(p) = \frac{p}{9} + 1-(1/3)^{p/3}$ to characterize the 2D sheet-like structure for incompressible MHD turbulence (referred to as MB model), respectively\footnote{We note that \citet{Politano1995} also derived a mathematically similar model, $Y(p) = p/8 + 1 - 1/2^{p/4}$, within the framework of Iroshnikov-Kraichnan isotropic turbulence (\citealt{Iroshnikov1963, Kraichnan1965}). However, the MB model is suitable within the context of the strong, anisotropic MHD turbulence described by GS95, which aligns with the focus of our study.}.

Based on the extended self-similarity hypothesis (\cite{Benzi1993} suggested that the power-law scaling can extend from the inertial range to the dissipative region), we adopt the scale exponent $Y(p)$ between the third-order and $p$-order structure functions to differentiate the level of intermittency (see the inset of Fig. \ref{fig_intermittency}(a) for details and \cite{Wang2024} for application to observations). Fig. \ref{fig_intermittency} shows the scaling exponent for the magnetic field as a function of the order in the cases of strong and weak turbulence, including the total $Y(p)$, parallel $Y_{\parallel}(p)$ and perpendicular $Y_{\perp}(p)$ components. As we can see, for the strong turbulence, the distribution of $Y(p)$ aligns closely with the MB model (panel (a)), which indicates the occurrence of stronger intermittency with the 2D sheet-like structure of magnetic fields. For the weak turbulence (see panels (b) and (c)), the distribution of $Y(p)$ is close to the SL model (with slightly weaker intermittency than strong turbulence), indicating the occurrence of the 1D vortex filament of magnetic fields\footnote{Given that sampling all the numbers $1024^3$ in a box is very time-consuming, we use a large step $\Delta R=64$ pixels to reduce the calculation time of the structure function. Hence, the slightly non-systematic trend of $Y_{\parallel}(p)$ and $Y_{\perp}(p)$ with $\chi$ is from incomplete statistics.}. It may be due to the strong turbulence enhancing magnetic fluctuations, leading to the localization of the energy cascade. For all $\chi$, the separated distributions of $Y_{\parallel}(p)$ and $Y_{\perp}(p)$ indicates the anisotropy of the intermittency, with $Y_{\perp}(p)$ stronger intermittency than $Y_{\parallel}(p)$ and $Y(p)$. It demonstrates that the dominated cascades in the perpendicular direction with respect to the local magnetic field enhance the behavior of intermittency.

\subsubsection{Spectra and Anisotropies of Generated Alfv\'en and Fast Modes}\label{sec:numre.prop.Palsma-modes}
\begin{figure}
\centering
\includegraphics[width=0.9\columnwidth,height=0.35\textheight]{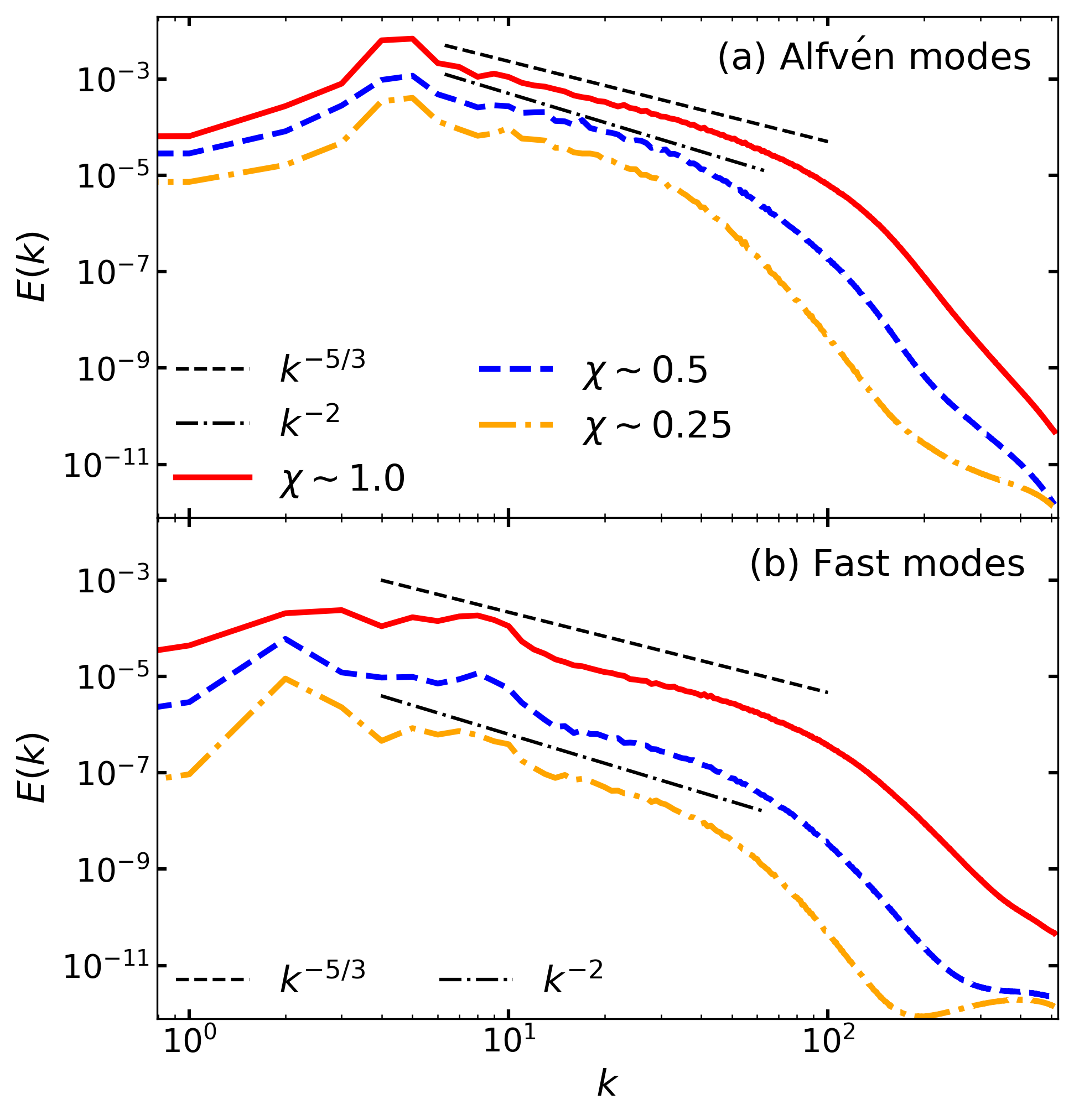}
\caption{The power spectra of the magnetic field for the Alfv\'en (panel (a)) and fast (panel (b)) modes. 
}
\label{fig_ps-af}
\end{figure}

In this section, we investigate the properties of the generated Alfv\'en and fast modes due to Alfv\'en-Alfv\'en wave interactions. The magnetic field power spectra of Alfv\'en modes are shown in Fig. \ref{fig_ps-af}(a), which remain a similar power-law relationship with the overall spectra for the strong and weak turbulence (see Fig. \ref{fig_ps}(a) for details). For the strong turbulence ($\chi \sim 1.0$), we have $E_{\rm A}(k)\propto k^{-5/3}$, consistent with the numerical results of the NRMHD turbulence (\citealt{Cho2002, Cho2003}). For the weak turbulence ($\chi \sim 0.5$ and 0.25), we have $E_{\rm A}(k)\propto k^{-2}$, consistent with the prediction of \cite{Galtier2000} in the non-relativistic case. The magnetic field power spectra of fast modes exhibit similar spectra to Alfv\'en modes but peak at much smaller $k$, as shown in Fig. \ref{fig_ps-af}(b), which is because they result from the nonlinear interactions of Alfv\'en modes. Notably, there is a low ratio of fast to Alfv\'en magnetic energy: (1) for $\chi \sim 1.0$, our results are similar to that of Cho05, which found that the ratio is roughly 0.13–0.15; (2) for $\chi \sim 0.5$ and 0.25, we find that the ratio is much smaller than that for $\chi \sim 1.0$, and the smaller $\chi$ the smaller ratio.

\begin{figure}
\centering
\includegraphics[width=0.9\columnwidth,height=0.3\textheight]{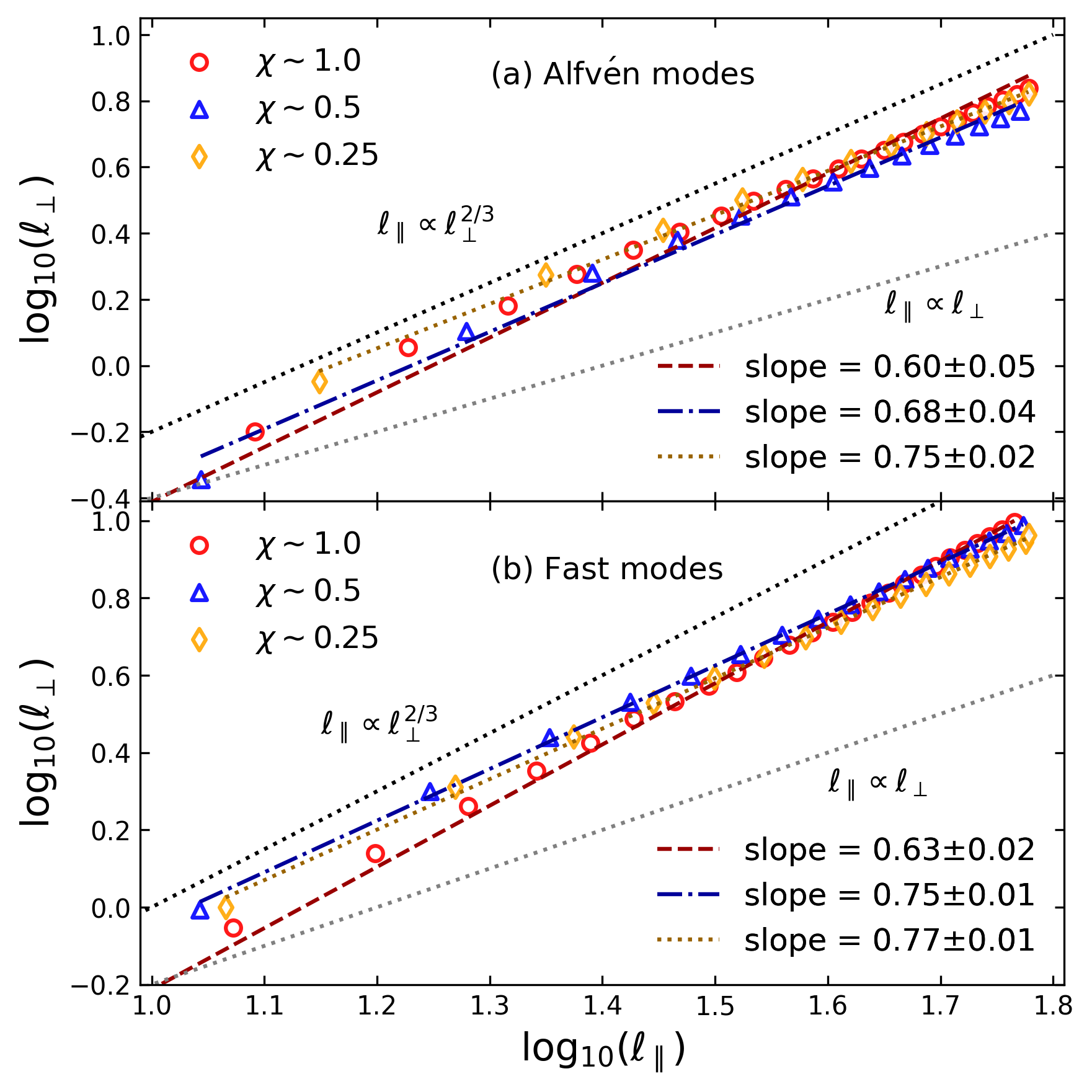}
\caption{Scale-dependent anisotropy of magnetic field for the Alfv\'en (panel (a)) and fast (panel (b)) modes. The coordinates $\ell_{\perp}$ and $\ell_{\parallel}$ represent the perpendicular and parallel scales of the eddies along the local magnetic field, respectively. The colored lines are the corresponding linear fitting.
}
\label{fig_anis-af}
\end{figure}

In addition, we also explore the scale-dependent anisotropy of Alfv\'en and fast modes on small scales in Fig. \ref{fig_anis-af}. For the strong turbulence ($\chi \sim 1$), Alfv\'en and fast modes present the scale-dependent anisotropic relation of $\ell_{\parallel} \propto \ell_{\perp}^{2/3}$. As the turbulent cascade weakens (decreasing the $\chi$ value), their scale-dependent anisotropies become weaker and close to $\ell_{\parallel} \propto \ell_{\perp}^{3/4}$, implying that the eddy distortions depend more weakly on the spatial scale (see also Fig. 6 of Cho05 for a contour map).

\section{Application to the Magnetospheres of Neutron Stars}\label{sec:application-NS}
In this section, we apply the relativistic force-free Alfv\'enic turbulence to the magnetospheres of neutron stars to understand their energy transfer and acceleration mechanisms. The Alfv\'en waves injected into the magnetosphere of a neutron star either damp near the surface of the neutron star or escape from the closed magnetic field line region, thereby driving relativistic outflows. Waves trapped near the surface of a star can undergo turbulent cascades. Taking Vela pulsar as an example, we estimate the Alfv\'en wave luminosity $L_{\rm A}$, rotational energy loss rate $L_{\rm sd}$, and total cascade luminosity $L_{\rm cas}$ via the following formulae (see TB98)
\begin{equation}
L_{\rm A} \sim 0.2 \left( \frac{\delta B_\star}{B_\star} \right)^{8/3} B_\star^2 R_\star^2 c, \label{eq.LA}
\end{equation}
\begin{equation}
L_{\rm sd} \sim 0.1 L_A (\Omega R_A / c)^2, \label{eq.Lsd}
\end{equation}
\begin{equation}
L_{\rm cas} \sim \frac{1}{6} \left( \frac{\delta B_\star}{B_\star} \right)^{4} \left( \frac{k_\perp}{k_\parallel} \right)^{2} \left( \frac{\nu_0 R_\star}{c} \right) B_\star^2 R_\star^2 c, \label{eq.Lcas}
\end{equation}
where $B_\star$ and $\delta B_\star$ are the magnetic field strength and its disturbance at the star surface (with the radius $R_\star$, the spin frequency $\Omega = 2\pi/P$, and the spin period $P$), respectively. Consider the Alfv\'en radius $R_{\rm A}$ equal to the outer scale of turbulence, where the Alfv\'en wave frequency is $\nu_0 = k_{\parallel} v_{\rm A}/ (2\pi)$ related to the parallel wavenumber $k_\parallel$ and the Alfv\'en velocity $v_{\rm A}$. Inserting the scale-dependent anisotropic relation of $k_{\parallel} \propto k_{\perp}^{2/3}$ confirmed numerically (see Fig. \ref{fig_anisotropy}) and $k=\sqrt{k_{\perp}^2 + k_{\parallel}^2} \sim k_\perp$ into Eq. (\ref{eq.Lcas}), we can obtain the cascade luminosity of
\begin{equation}
L_{\rm cas} \sim \frac{1}{12\pi} \left(\frac{\delta B_\star}{B_\star}\right)^4 k^{4/3} B_\star^2 R_\star^2 c. \label{eq.Lcas-k}
\end{equation}

\begin{figure}
\centering
\includegraphics[width=0.99\columnwidth,height=0.40\textheight]{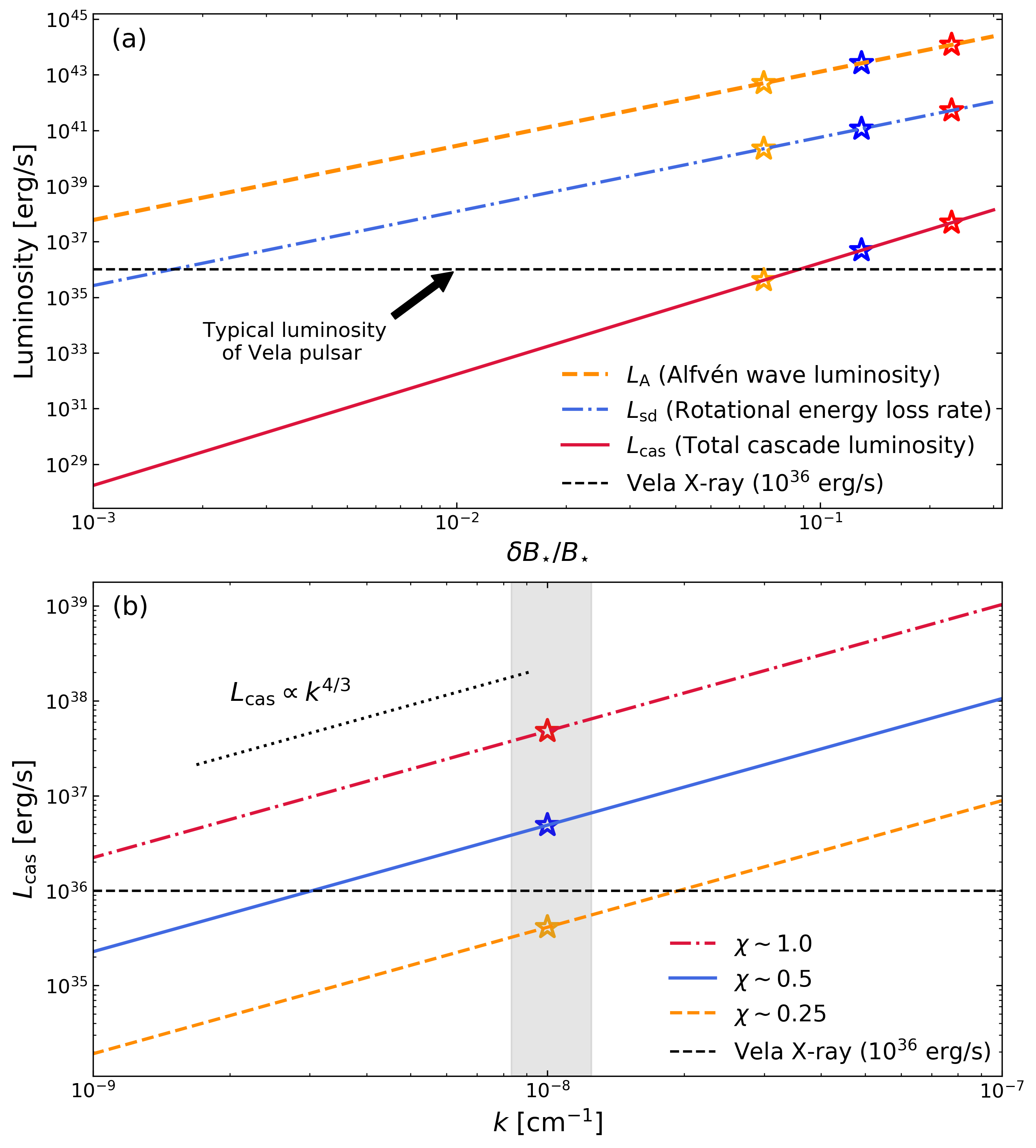}
\caption{Luminosity vs. the amplitude of magnetic field disturbance (panel (a)) and the wavenumber (panel (b)). The symbols plotted by the red, blue, and orange stars correspond to $\chi \sim 1.0$ ($\delta B_{\star}/B_{\star}=0.23$), 0.5 (0.13), and 0.25 (0.07), respectively. The vertically shaded band in panel (b) represents the outer scale of $L_{\rm outer} \simeq (1.0\pm 0.2)\ R_{\rm A}$.
}
\label{fig_luminosity}
\end{figure}

Adopting some typical parameter values of Vela pulsar: $B_\star \sim 10^{12}$ G, $R_\star \sim 10$ km, $R_{\rm A} \sim 100 R_\star = 10^3$ km, $k_{\perp} \sim 1/R_{\rm A}$ and $P \sim 0.1$ s (\citealt{Goldreich1969, Manchester2005, Lattimer2007, Carli2024}), we show the relation of luminosities with the amplitude of magnetic field disturbance in Fig. \ref{fig_luminosity}(a). As seen, the different luminosities increase with $\delta B_{\star}/B_{\star}$, indicating that the stronger the turbulence is, the higher the luminosities. The Alfv\'en luminosity $L_{\rm A}$ and the spin-down luminosity $L_{\rm sd}$ are greater than the typical luminosity of Vela ($10^{36}$ erg s$^{-1}$) observed at X-ray bands. For the strong ($\chi \sim 1.0$) and weak ($\chi \sim 0.5$, and 0.25) turbulence, we have the cascade luminosities of $L_{\rm cas} = 48,\ 4.9,\ 0.412 \times 10^{36}$ erg s$^{-1}$. In the case of strong (or moderately weak $\chi \sim 0.5$) turbulence, we thus expect that the turbulent cascade via accelerating particles can reproduce the radiative luminosity of the Vela pulsar. 

As for the strong and weak turbulence, we plot the relation that we found between the cascade luminosity and the wavenumber in Fig. \ref{fig_luminosity}(b) (see also Eq. (\ref{eq.Lcas-k})). As shown, the cascade luminosity decreases with the spatial scale ($\sim 1/k$). In the case of weak turbulence (such as $\chi \sim 0.25$), the turbulent cascade can only provide the expected luminosity ($> 10^{36}$ erg s$^{-1}$) close to the neutron star surface, i.e., inner magnetosphere, where the produced emission may be attenuated due to electromagnetic cascade processes and cannot reach to observers. Therefore, we expect that the strong (or moderately weak $\chi \sim 0.5$) turbulent cascade can explain the X-ray radiation of the Vela pulsar. Nevertheless, we would like to mention to the interested reader that the turbulent correlation scale is of great significance to determine the specific luminosity of the turbulent cascade. Besides, it needs to consider the geometry of the open magnetic field lines of the neutron stars. Our current work offers a limited perspective on the application of relativistic force-free MHD turbulence.

\section{Discussion}\label{sec:discu}
Under the condition of relativistic force-free MHD turbulence, the Alfv\'en-Alfv\'en wave interactions are the dominant wave-wave interactions (see TB98). Therefore, we only explored the turbulent cascade between two Alfv\'en waves. However, the fast-fast and Alfv\'en-fast wave interactions driving may be important as well. For instance, the former plays a crucial role in particle acceleration and magnetic field amplification in environments such as supernova remnants (\citealt{Yan2004}) and intracluster media (\citealt{Brunetti2007}). The latter is essential for magnetic reconnection and plasma heating in the solar wind (\citealt{Verdini2018, Shoda2019}). We plan to explore these turbulence properties in future work.  

Although we only provided the numerical results for the plasma parameters $\beta=0.2$ in this work, we would like to mention to the interested reader that the influence of $\beta$ on turbulence properties is negligible in the case of strong turbulence (given the computational resource overhead, we did not test for the weak turbulence). The main reason why force-free MHD turbulence is insensitive to $\beta$ lies in its fundamental condition that magnetic tension completely dominates dynamics and the gas pressure effect is ignored (TB98). 

To better characterize anisotropy in weak turbulence, we combine the second-order structure function and the power spectrum. The former shows that weak turbulence exhibits a scale-dependent anisotropy approximately $k_{\parallel} \propto k_{\perp}^{2/3}$ at small scales (see Fig. \ref{fig_anisotropy}, similar to the results of \cite{Ng2007}), and the latter exhibits the theoretically expected relation of $k_{\parallel} \simeq const$ on large scales. This paper focuses here on the small-scale results from the structure function, as it has a limited number of samples on large scales, leading to significant statistical errors, which is a limitation of the two-point correlation function. The second-order power spectrum, transformed from the position space to the wavenumber space, may also affect quantitative spatial anisotropy, particularly at small scales where sampling is sparse and statistical reliability is lower (see Fig. \ref{fig_pdf}, the non-Gaussian distribution of the magnetic field PDF indicates that the second-order power spectrum and structure function cannot fully describe anisotropy when the ``inertial" range is too short, as shown in Fig. \ref{fig_ps}). Future studies incorporating higher-order statistics, such as the bispectrum and trispectrum, may yield more self-consistent physical insight.

By analyzing the relation of $\chi$ vs. $k_{\perp}$ in the weak turbulence, we found that $\chi$ increases in a way of $\chi \propto k_{\perp}^{1/2}$ in the inertial range, which are similar to the previous simulation results (\citealt{Ng2007,Ripperda2021}). As described in Section \ref{sec:numre.evolu}, the evolution timescale of weak turbulence to fully developed turbulence is much longer than that of the strong one; that is, the weaker the turbulence, the longer the interaction timescale (see \cite{Galtier2000} for a theoretical prediction). This can probably be attributed to two primary reasons. One can be attributed to the strength of nonlinear interaction. The nonlinear interaction in weak turbulence is weaker than that in the strong one; that is, the weaker the nonlinear effect, the slower the inter-wave energy exchange. Another one is related to the energy transfer mechanism. In weak turbulence, energy transfers mainly by four-wave resonant interactions (see TB98; \citealt{Saur2002}; accompanied by a weak nonlinear process), reducing the transformation rate (\citealt{Galtier2000}). While in strong turbulence, energy transfers effectively without a stringent resonant condition. 

We carried out simulations of RMHD turbulence in the force-free approximation, a pivotal adoption that defines the theoretical model, numerical approach, and applicable astrophysical domains for RMHD turbulence. Besides the differences in physical assumptions and numerical implementation, a fundamental distinction between force-free and fully relativistic MHD lies in their energy transfer and dissipation mechanisms: the former cascades energy primarily via Alfv\'en/fast mode interactions within the electromagnetic field, while the latter involves multi-channel conversions between magnetic, kinetic, and thermal energy, leading to richer dissipation processes. The force-free model accurately captures MHD turbulence properties in magnetically dominated environments. Still, its applicability is limited in high-gas-pressure scenarios, where full relativistic MHD is required (e.g., \citealt{Zrake2012}). Here, we mention that the force-free approximation has its intrinsic limitations. It cannot describe how energy is dissipated on kinetic scales due to the neglect of matter inertia (see Section 3 in TB98). It still requires a sufficient number of charge carriers to maintain the current density and ensure that $\mathbf{E} \cdot \mathbf{B} = 0$. If the density of the background plasma is too low to provide the current required for cascading, ``charge starvation" will occur. In this context, a strong parallel electric field accelerates particles, compromising the force-free approximation, which can no longer be sustained (see Section 7 in TB98).

Our decaying turbulence simulation is initialized by exciting Alfv\'en wave perturbations (see Eq. (\ref{eq.drive-k})) rather than driving an external force. The decay time is estimated by $\tau_{\rm decay} \sim \tau_{\rm w}/\chi^2 \sim 4/\chi^2$ with the wave-crossing time $\tau_{\rm w} \sim \ell_\parallel/c \sim 2\pi/1.5$, approximately corresponding to 4, 16, and 64 in units of the code for $\chi=1.0$, 0.5, and 0.25, respectively. As seen in Fig. \ref{fig_pst}, most of the magnetic energy remains concentrated near the initial wavenumber with an approximately constant length scale. We analyzed the decay of magnetic energy over time using the fluctuating magnetic energy $b^2$, as the total magnetic energy is dominated by the uniform background field $B_0$ in our high-magnetization setup. The measured decay exponent is $\sim$0.3 for $b^2$ and $\sim$0.4 for the total fluctuating energy $b^2+V^2$. These values are lower than the exponents (e.g., $2/3$ or $10/9$) predicted for non-relativistic MHD turbulence decay based on conserved quantities like net magnetic helicity or the Hosking integral. These results align with key findings in the simulation (\citealt{Zrake2016}) and theoretical (\citealt{Hosking2021}) work. This discrepancy may be explained by the specific physics of our force-free system: energy is confined solely to the magnetic field, removing an independent kinetic energy cascade channel. Additionally, in our high-resolution setup with a high effective Lundquist number, the instantaneous decay rate can be suppressed relative to the Alfvén rate, leading to a shallower effective exponent over the finite simulation time.

It is noteworthy that our decaying turbulence simulations do not exhibit an inverse energy cascade. This absence can be attributed to our initial condition of zero net magnetic helicity and to the force-free framework, where energy is confined almost to the electromagnetic field, and the coupling between kinetic and magnetic energy is severely weakened (decoupled). The predominance of a forward cascade aligns with prior analytical and numerical studies of Alfv\'enic turbulence in the force-free regime (e.g., \citealt{Thompson1998, Cho2005}).

Although we did not quantify magnetic helicity or its fluctuations in this paper, significant local magnetic helicity fluctuations could still exist despite $\mathbf{E} \cdot \mathbf{B} = 0$. Recent work by \cite{Hosking2021, Hosking2023} has established that the Hosking integral, which characterizes magnetic helicity fluctuations in subvolumes, serves as a key conserved quantity in decaying turbulence with zero net magnetic helicity, governing the decay process. Therefore, applying the Hosking integral framework to our force-free MHD turbulence simulations represents an interesting topic for future research.

The study of RMHD turbulence is crucial for understanding energy transfer, magnetic field evolution, and particle acceleration in extreme astrophysical environments, such as the magnetospheres of neutron stars and the accretion disks of black holes. The turbulent cascade luminosity is associated with the turbulent magnetic field amplitude and the spatial scales, which can help understand pulsar radio emission, X-ray flares, and even electromagnetic precursors to gravitational wave events. Future research should combine high-resolution simulations with multi-messenger observations (e.g., IXPE, FAST) to refine turbulence models. Investigating RMHD turbulence is expected to unveil new energy cascade paradigms under extreme magnetic fields, potentially leading to a unified framework for understanding magnetic field evolution in compact objects and the formation of gravitational wave sources.

\section{Summary}\label{sec:summa}
In this work, we first simulated strong and weak relativistic force-free MHD turbulence at a high numerical resolution of $1024^3$, exploring fundamental turbulence properties such as the energy spectrum, anisotropy, intermittency, and characteristics of the generated Alfv\'en and fast modes. We then applied our numerical results to neutron star magnetospheres to understand their energy transfer and acceleration mechanisms. Our results are summarized as follows:

\begin{enumerate}
\item We find that the power spectra of the magnetic field follow power-law relationships of $E(k) \propto k^{-5/3}$ for strong turbulence and $E(k) \propto k^{-2}$ for weak turbulence, which align with previous theoretical and numerical studies in the NRMHD turbulence.

\item The energy cascade perpendicular to the local magnetic field direction dominates the energy transfer process of the relativistic force-free MHD turbulence. The strong turbulence exhibits a scale-dependent anisotropic scaling of $\ell_{\parallel} \propto \ell_{\perp}^{2/3}$, which also holds for the weak turbulence on small scales.    

\item Strong turbulence exhibits elongated non-Gaussian tails of the magnetic field PDF, corresponding to a stronger intermittency closer to the MB model, while weak turbulence without significant tails shows a weaker intermittency closer to the SL model. Their intermittencies are mainly dominated by the structure perpendicular to the local magnetic field direction.

\item The Alfv\'en and fast modes exhibit an energy spectrum of $E(k) \propto k^{-\alpha}$, where $\alpha = 5/3$ and $2$ for strong and weak turbulence, respectively, and a scale-dependent anisotropic relation of $\ell_{\parallel} \propto \ell_{\perp}^{\gamma}$, where $\gamma \simeq 2/3$ and $3/4$ for strong and weak turbulence, respectively. 

\item The turbulent cascade luminosity increases with the wavenumber in the way of a power-law relationship of $L_{\rm cas} \propto k^{4/3}$, and the strong (or moderately weak $\chi \sim 0.5$) turbulent cascade can explain the X-ray radiation of the Vela pulsar. 

\end{enumerate}

\begin{acknowledgments}
We sincerely thank the anonymous referee for constructive comments that significantly improved our manuscript. This work is supported by the high-performance computing platform of the School of Physics and Optoelectronics, Xiangtan University. The authors thank the support from the National Natural Science Foundation of China (grant No. 11973035). J.F.Z. also thanks the Hunan Natural Science Foundation for Distinguished Young Scholars (No. 2023JJ10039). J.C. is supported by the Korea Astronomy and Space Science Institute under the R\&D program (Project No. 2025-9-844-00) supervised by the Korea AeroSpace Administration, and the Yunnan Provincial Foreign Talent Introduction Program (No. 202505AP120035). N.N.G is grateful for the support from the Xiangtan University Innovation Foundation for Post-graduate (No. XDCX2025Y257).
\end{acknowledgments}

\bibliography{ms}{}

@ARTICLE{Benzi1993,
       author = {{Benzi}, R. and {Ciliberto}, S. and {Tripiccione}, R. and {Baudet}, C. and {Massaioli}, F. and {Succi}, S.},
        title = "{Extended self-similarity in turbulent flows}",
      journal = {\pre},
         year = 1993,
        month = jul,
       volume = {48},
       number = {1},
        pages = {R29-R32},
          doi = {10.1103/PhysRevE.48.R29},
       adsurl = {https://ui.adsabs.harvard.edu/abs/1993PhRvE..48...29B}
}

@BOOK{Beresnyak2019,
       author = {{Beresnyak}, Andrey and {Lazarian}, Alexander},
        title = "{Turbulence in Magnetohydrodynamics}",
         year = 2019,
       adsurl = {https://ui.adsabs.harvard.edu/abs/2019tuma.book.....B}
}

@ARTICLE{Blandford1977,
       author = {{Blandford}, R.~D. and {Znajek}, R.~L.},
        title = "{Electromagnetic extraction of energy from Kerr black holes.}",
      journal = {\mnras},
         year = 1977,
        month = may,
       volume = {179},
        pages = {433-456},
          doi = {10.1093/mnras/179.3.433},
       adsurl = {https://ui.adsabs.harvard.edu/abs/1977MNRAS.179..433B}
}

@INPROCEEDINGS{Boldyrev2012,
       author = {{Boldyrev}, Stanislav and {Perez}, Jean Carlos and {Zhdankin}, Vladimir},
        title = "{Residual energy in MHD turbulence and in the solar wind}",
    booktitle = {Physics of the Heliosphere: A 10 Year Retrospective},
         year = 2012,
       editor = {{Heerikhuisen}, Jacob and {Li}, Gang and {Pogorelov}, Nikolai and {Zank}, Gary},
       series = {American Institute of Physics Conference Series},
       volume = {1436},
        month = may,
        pages = {18-23},
          doi = {10.1063/1.4723584},
archivePrefix = {arXiv},
       eprint = {1108.6072},
 primaryClass = {astro-ph.GA},
       adsurl = {https://ui.adsabs.harvard.edu/abs/2012AIPC.1436...18B}
}

@ARTICLE{Brunetti2007,
       author = {{Brunetti}, G. and {Lazarian}, A.},
        title = "{Compressible turbulence in galaxy clusters: physics and stochastic particle re-acceleration}",
      journal = {\mnras},
         year = 2007,
        month = jun,
       volume = {378},
       number = {1},
        pages = {245-275},
          doi = {10.1111/j.1365-2966.2007.11771.x},
archivePrefix = {arXiv},
       eprint = {astro-ph/0703591},
 primaryClass = {astro-ph},
       adsurl = {https://ui.adsabs.harvard.edu/abs/2007MNRAS.378..245B}
}

@ARTICLE{Carli2024,
       author = {{Carli}, E. and {Antonopoulou}, D. and {Burgay}, M. and {Keith}, M.~J. and {Levin}, L. and {Liu}, Y. and {Stappers}, B.~W. and {Turner}, J.~D. and {Barr}, E.~D. and {Breton}, R.~P. and {Buchner}, S. and {Kramer}, M. and {Padmanabh}, P.~V. and {Possenti}, A. and {Venkatraman Krishnan}, V. and {Venter}, C. and {Becker}, W. and {Maitra}, C. and {Haberl}, F. and {Thongmeearkom}, T.},
        title = "{The TRAPUM Small Magellanic Cloud pulsar survey with MeerKAT - II. Nine new radio timing solutions and glitches from young pulsars}",
      journal = {\mnras},
         year = 2024,
        month = oct,
       volume = {533},
       number = {4},
        pages = {3957-3974},
          doi = {10.1093/mnras/stae1897},
archivePrefix = {arXiv},
       eprint = {2408.01965},
 primaryClass = {astro-ph.HE},
       adsurl = {https://ui.adsabs.harvard.edu/abs/2024MNRAS.533.3957C}
}

@ARTICLE{Chandran2015,
       author = {{Chandran}, B.~D.~G. and {Schekochihin}, A.~A. and {Mallet}, A.},
        title = "{Intermittency and Alignment in Strong RMHD Turbulence}",
      journal = {\apj},
         year = 2015,
        month = jul,
       volume = {807},
       number = {1},
          eid = {39},
        pages = {39},
          doi = {10.1088/0004-637X/807/1/39},
archivePrefix = {arXiv},
       eprint = {1403.6354},
 primaryClass = {astro-ph.SR},
       adsurl = {https://ui.adsabs.harvard.edu/abs/2015ApJ...807...39C}
}

@ARTICLE{Cho2002,
       author = {{Cho}, Jungyeon and {Lazarian}, A.},
        title = "{Compressible Sub-Alfv{\'e}nic MHD Turbulence in Low- {\ensuremath{\beta}} Plasmas}",
      journal = {\prl},
         year = 2002,
        month = jun,
       volume = {88},
       number = {24},
          eid = {245001},
        pages = {245001},
          doi = {10.1103/PhysRevLett.88.245001},
archivePrefix = {arXiv},
       eprint = {astro-ph/0205282},
 primaryClass = {astro-ph},
       adsurl = {https://ui.adsabs.harvard.edu/abs/2002PhRvL..88x5001C}
}

@ARTICLE{Cho2003,
       author = {{Cho}, Jungyeon and {Lazarian}, A.},
        title = "{Compressible magnetohydrodynamic turbulence: mode coupling, scaling relations, anisotropy, viscosity-damped regime and astrophysical implications}",
      journal = {\mnras},
         year = 2003,
        month = oct,
       volume = {345},
       number = {12},
        pages = {325-339},
          doi = {10.1046/j.1365-8711.2003.06941.x},
archivePrefix = {arXiv},
       eprint = {astro-ph/0301062},
 primaryClass = {astro-ph},
       adsurl = {https://ui.adsabs.harvard.edu/abs/2003MNRAS.345..325C}
}

@ARTICLE{Cho2003b,
       author = {{Cho}, Jungyeon and {Lazarian}, A. and {Vishniac}, Ethan T.},
        title = "{Ordinary and Viscosity-damped Magnetohydrodynamic Turbulence}",
      journal = {\apj},
         year = 2003,
        month = oct,
       volume = {595},
       number = {2},
        pages = {812-823},
          doi = {10.1086/377515},
archivePrefix = {arXiv},
       eprint = {astro-ph/0305212},
 primaryClass = {astro-ph},
       adsurl = {https://ui.adsabs.harvard.edu/abs/2003ApJ...595..812C}
}

@ARTICLE{Cho2004,
       author = {{Cho}, Jungyeon and {Lazarian}, A.},
        title = "{The Anisotropy of Electron Magnetohydrodynamic Turbulence}",
      journal = {\apjl},
         year = 2004,
        month = nov,
       volume = {615},
       number = {1},
        pages = {L41-L44},
          doi = {10.1086/425215},
archivePrefix = {arXiv},
       eprint = {astro-ph/0406595},
 primaryClass = {astro-ph},
       adsurl = {https://ui.adsabs.harvard.edu/abs/2004ApJ...615L..41C}
}

@ARTICLE{Cho2005,
       author = {{Cho}, Jungyeon},
        title = "{Simulations of Relativistic Force-free Magnetohydrodynamic Turbulence}",
      journal = {\apj},
         year = 2005,
        month = mar,
       volume = {621},
       number = {1},
        pages = {324-327},
          doi = {10.1086/427493},
archivePrefix = {arXiv},
       eprint = {astro-ph/0408318},
 primaryClass = {astro-ph},
       adsurl = {https://ui.adsabs.harvard.edu/abs/2005ApJ...621..324C}
}

@ARTICLE{Cho2000,
       author = {{Cho}, Jungyeon and {Vishniac}, Ethan T.},
        title = "{The Anisotropy of Magnetohydrodynamic Alfv{\'e}nic Turbulence}",
      journal = {\apj},
         year = 2000,
        month = aug,
       volume = {539},
       number = {1},
        pages = {273-282},
          doi = {10.1086/309213},
archivePrefix = {arXiv},
       eprint = {astro-ph/0003403},
 primaryClass = {astro-ph},
       adsurl = {https://ui.adsabs.harvard.edu/abs/2000ApJ...539..273C}
}

@ARTICLE{Cho2014,
       author = {{Cho}, Jungyeon and {Lazarian}, A.},
        title = "{Imbalanced Relativistic Force-free Magnetohydrodynamic Turbulence}",
      journal = {\apj},
         year = 2014,
        month = jan,
       volume = {780},
       number = {1},
          eid = {30},
        pages = {30},
          doi = {10.1088/0004-637X/780/1/30},
archivePrefix = {arXiv},
       eprint = {1312.6128},
 primaryClass = {astro-ph.HE},
       adsurl = {https://ui.adsabs.harvard.edu/abs/2014ApJ...780...30C}
}

@ARTICLE{Duncan1992,
       author = {{Duncan}, Robert C. and {Thompson}, Christopher},
        title = "{Formation of Very Strongly Magnetized Neutron Stars: Implications for Gamma-Ray Bursts}",
      journal = {\apjl},
         year = 1992,
        month = jun,
       volume = {392},
        pages = {L9},
          doi = {10.1086/186413},
       adsurl = {https://ui.adsabs.harvard.edu/abs/1992ApJ...392L...9D}
}

@ARTICLE{Gao2024,
       author = {{Gao}, Na-Na and {Zhang}, Jian-Fu},
        title = "{The Diffusion and Scattering of Accelerating Particles in Compressible MHD Turbulence}",
      journal = {\apj},
         year = 2024,
        month = jan,
       volume = {961},
       number = {1},
          eid = {80},
        pages = {80},
          doi = {10.3847/1538-4357/ad0d9e},
archivePrefix = {arXiv},
       eprint = {2311.09554},
 primaryClass = {astro-ph.HE},
       adsurl = {https://ui.adsabs.harvard.edu/abs/2024ApJ...961...80G}
}

@ARTICLE{Gao2025,
       author = {{Gao}, Na-Na and {Zhang}, Jian-Fu},
        title = "{Anisotropic diffusion of high-energy cosmic rays in magnetohydrodynamic turbulence}",
      journal = {\aap},
         year = 2025,
        month = feb,
       volume = {694},
          eid = {A201},
        pages = {A201},
          doi = {10.1051/0004-6361/202452541},
archivePrefix = {arXiv},
       eprint = {2501.04986},
 primaryClass = {astro-ph.HE},
       adsurl = {https://ui.adsabs.harvard.edu/abs/2025A&A...694A.201G}
}

@ARTICLE{Galtier2000,
       author = {{Galtier}, S. and {Nazarenko}, S.~V. and {Newell}, A.~C. and {Pouquet}, A.},
        title = "{A weak turbulence theory for incompressible magnetohydrodynamics}",
      journal = {Journal of Plasma Physics},
         year = 2000,
        month = jun,
       volume = {63},
       number = {5},
        pages = {447-488},
          doi = {10.1017/S0022377899008284},
archivePrefix = {arXiv},
       eprint = {astro-ph/0008148},
 primaryClass = {astro-ph},
       adsurl = {https://ui.adsabs.harvard.edu/abs/2000JPlPh..63..447G}
}

@ARTICLE{Goldreich1969,
       author = {{Goldreich}, Peter and {Julian}, William H.},
        title = "{Pulsar Electrodynamics}",
      journal = {\apj},
         year = 1969,
        month = aug,
       volume = {157},
        pages = {869},
          doi = {10.1086/150119},
       adsurl = {https://ui.adsabs.harvard.edu/abs/1969ApJ...157..869G}
}

@ARTICLE{Goldreich1995,
       author = {{Goldreich}, P. and {Sridhar}, S.},
        title = "{Toward a Theory of Interstellar Turbulence. II. Strong Alfvenic Turbulence}",
      journal = {\apj},
         year = 1995,
        month = jan,
       volume = {438},
        pages = {763},
          doi = {10.1086/175121},
       adsurl = {https://ui.adsabs.harvard.edu/abs/1995ApJ...438..763G}
}

@ARTICLE{Harten1983,
       author = {Harten, Amiram and Lax, Peter D. and Leer, Bram van},
        title = "{On Upstream Differencing and Godunov-Type Schemes for Hyperbolic Conservation Laws}",
      journal = {SIAM Review},
         year = 1983,
       volume = {25},
        pages = {35-61},
       doi = {10.1137/1025002},
       URL = {https://doi.org/10.1137/1025002}
}

@ARTICLE{Hosking2021,
       author = {{Hosking}, David N. and {Schekochihin}, Alexander A.},
        title = "{Reconnection-Controlled Decay of Magnetohydrodynamic Turbulence and the Role of Invariants}",
      journal = {Physical Review X},
         year = 2021,
        month = oct,
       volume = {11},
       number = {4},
          eid = {041005},
        pages = {041005},
          doi = {10.1103/PhysRevX.11.041005},
archivePrefix = {arXiv},
       eprint = {2012.01393},
 primaryClass = {physics.flu-dyn},
       adsurl = {https://ui.adsabs.harvard.edu/abs/2021PhRvX..11d1005H}
}

@ARTICLE{Hosking2023,
       author = {{Hosking}, David N. and {Schekochihin}, Alexander A.},
        title = "{Cosmic-void observations reconciled with primordial magnetogenesis}",
      journal = {Nature Communications},
         year = 2023,
        month = nov,
       volume = {14},
          eid = {7523},
        pages = {7523},
          doi = {10.1038/s41467-023-43258-3},
archivePrefix = {arXiv},
       eprint = {2203.03573},
 primaryClass = {astro-ph.CO},
       adsurl = {https://ui.adsabs.harvard.edu/abs/2023NatCo..14.7523H}
}

@ARTICLE{Iroshnikov1963,
       author = {{Iroshnikov}, P.~S.},
        title = "{Turbulence of a Conducting Fluid in a Strong Magnetic Field}",
      journal = {\azh},
         year = 1963,
        month = jan,
       volume = {40},
        pages = {742},
       adsurl = {https://ui.adsabs.harvard.edu/abs/1963AZh....40..742I}
}

@ARTICLE{Kolmogorov1941,
       author = {{Kolmogorov}, A.},
        title = "{The Local Structure of Turbulence in Incompressible Viscous Fluid for Very Large Reynolds' Numbers}",
      journal = {Akademiia Nauk SSSR Doklady},
         year = 1941,
        month = jan,
       volume = {30},
        pages = {301-305},
       adsurl = {https://ui.adsabs.harvard.edu/abs/1941DoSSR..30..301K}
}

@ARTICLE{Komissarov2002,
       author = {{Komissarov}, S.~S.},
        title = "{Time-dependent, force-free, degenerate electrodynamics}",
      journal = {\mnras},
         year = 2002,
        month = nov,
       volume = {336},
       number = {3},
        pages = {759-766},
          doi = {10.1046/j.1365-8711.2002.05313.x},
archivePrefix = {arXiv},
       eprint = {astro-ph/0202447},
 primaryClass = {astro-ph},
       adsurl = {https://ui.adsabs.harvard.edu/abs/2002MNRAS.336..759K}
}

@ARTICLE{Kraichnan1965,
       author = {{Kraichnan}, Robert H.},
        title = "{Inertial-Range Spectrum of Hydromagnetic Turbulence}",
      journal = {Physics of Fluids},
         year = 1965,
        month = jul,
       volume = {8},
       number = {7},
        pages = {1385-1387},
          doi = {10.1063/1.1761412},
       adsurl = {https://ui.adsabs.harvard.edu/abs/1965PhFl....8.1385K}
}

@ARTICLE{Kurganov2001,
       author = {{Kurganov}, Alexander and {Noelle}, Sebastian and {Petrova}, Guergana},
        title = "{Semidiscrete Central-Upwind Schemes for Hyperbolic Conservation Laws and Hamilton-Jacobi Equations}",
      journal = {SIAM Journal on Scientific Computing},
         year = 2001,
        month = jan,
       volume = {23},
       number = {3},
        pages = {707-740},
          doi = {10.1137/S1064827500373413},
       adsurl = {https://ui.adsabs.harvard.edu/abs/2001SJSC...23..707K}
}

@ARTICLE{Kuznetsov2001,
       author = {{Kuznetsov}, E.~A.},
        title = "{Weak Magnetohydrodynamic Turbulence of a Magnetized Plasma}",
      journal = {Soviet Journal of Experimental and Theoretical Physics},
         year = 2001,
        month = nov,
       volume = {93},
       number = {5},
        pages = {1052-1064},
          doi = {10.1134/1.1427116},
archivePrefix = {arXiv},
       eprint = {physics/0109066},
 primaryClass = {physics.plasm-ph},
       adsurl = {https://ui.adsabs.harvard.edu/abs/2001JETP...93.1052K}
}

@ARTICLE{Lattimer2007,
       author = {{Lattimer}, James M. and {Prakash}, Madappa},
        title = "{Neutron star observations: Prognosis for equation of state constraints}",
      journal = {\physrep},
         year = 2007,
        month = apr,
       volume = {442},
       number = {1-6},
        pages = {109-165},
          doi = {10.1016/j.physrep.2007.02.003},
archivePrefix = {arXiv},
       eprint = {astro-ph/0612440},
 primaryClass = {astro-ph},
       adsurl = {https://ui.adsabs.harvard.edu/abs/2007PhR...442..109L}
}

@ARTICLE{Li2019,
       author = {{Li}, Xinyu and {Zrake}, Jonathan and {Beloborodov}, Andrei M.},
        title = "{Dissipation of Alfv{\'e}n Waves in Relativistic Magnetospheres of Magnetars}",
      journal = {\apj},
         year = 2019,
        month = aug,
       volume = {881},
       number = {1},
          eid = {13},
        pages = {13},
          doi = {10.3847/1538-4357/ab2a03},
archivePrefix = {arXiv},
       eprint = {1810.10493},
 primaryClass = {astro-ph.HE},
       adsurl = {https://ui.adsabs.harvard.edu/abs/2019ApJ...881...13L}
}

@ARTICLE{Liang2023,
       author = {{Liang}, Shi-Min and {Zhang}, Jian-Fu and {Gao}, Na-Na and {Xiao}, Hua-Ping},
        title = "{Magnetic-reconnection-driven Turbulence and Turbulent Reconnection Acceleration}",
      journal = {\apj},
         year = 2023,
        month = aug,
       volume = {952},
       number = {2},
          eid = {93},
        pages = {93},
          doi = {10.3847/1538-4357/acdc18},
archivePrefix = {arXiv},
       eprint = {2306.03418},
 primaryClass = {astro-ph.HE},
       adsurl = {https://ui.adsabs.harvard.edu/abs/2023ApJ...952...93L}
}

@ARTICLE{Liang2025,
       author = {{Liang}, Shi-Min and {Zhang}, Jian-Fu and {Gao}, Na-Na and {Yi}, Nian-Yu},
        title = "{Studying the properties of reconnection-driven turbulence}",
      journal = {arXiv e-prints},
         year = 2025,
        month = oct,
          eid = {arXiv:2510.09978},
        pages = {arXiv:2510.09978},
          doi = {10.48550/arXiv.2510.09978},
archivePrefix = {arXiv},
       eprint = {2510.09978},
 primaryClass = {astro-ph.HE},
       adsurl = {https://ui.adsabs.harvard.edu/abs/2025arXiv251009978L}
}

@ARTICLE{Lyutikov2003,
       author = {{Lyutikov}, Maxim and {Blandford}, Roger},
        title = "{Gamma Ray Bursts as Electromagnetic Outflows}",
      journal = {arXiv e-prints},
         year = 2003,
        month = dec,
          eid = {astro-ph/0312347},
        pages = {astro-ph/0312347},
          doi = {10.48550/arXiv.astro-ph/0312347},
archivePrefix = {arXiv},
       eprint = {astro-ph/0312347},
 primaryClass = {astro-ph},
       adsurl = {https://ui.adsabs.harvard.edu/abs/2003astro.ph.12347L}
}

@ARTICLE{Manchester2005,
       author = {{Manchester}, R.~N. and {Hobbs}, G.~B. and {Teoh}, A. and {Hobbs}, M.},
        title = "{The Australia Telescope National Facility Pulsar Catalogue}",
      journal = {\aj},
         year = 2005,
        month = apr,
       volume = {129},
       number = {4},
        pages = {1993-2006},
          doi = {10.1086/428488},
archivePrefix = {arXiv},
       eprint = {astro-ph/0412641},
 primaryClass = {astro-ph},
       adsurl = {https://ui.adsabs.harvard.edu/abs/2005AJ....129.1993M}
}

@ARTICLE{Maron2001,
       author = {{Maron}, Jason and {Goldreich}, Peter},
        title = "{Simulations of Incompressible Magnetohydrodynamic Turbulence}",
      journal = {\apj},
         year = 2001,
        month = jun,
       volume = {554},
       number = {2},
        pages = {1175-1196},
          doi = {10.1086/321413},
archivePrefix = {arXiv},
       eprint = {astro-ph/0012491},
 primaryClass = {astro-ph},
       adsurl = {https://ui.adsabs.harvard.edu/abs/2001ApJ...554.1175M}
}

@ARTICLE{Meyrand2015,
       author = {{Meyrand}, R. and {Kiyani}, K.~H. and {Galtier}, S.},
        title = "{Weak magnetohydrodynamic turbulence and intermittency}",
      journal = {Journal of Fluid Mechanics},
         year = 2015,
        month = may,
       volume = {770},
          eid = {R1},
        pages = {R1},
          doi = {10.1017/jfm.2015.141},
       adsurl = {https://ui.adsabs.harvard.edu/abs/2015JFM...770R...1M}
}

@ARTICLE{Meyrand2016,
       author = {{Meyrand}, Romain and {Galtier}, S{\'e}bastien and {Kiyani}, Khurom H.},
        title = "{Direct Evidence of the Transition from Weak to Strong Magnetohydrodynamic Turbulence}",
      journal = {\prl},
         year = 2016,
        month = mar,
       volume = {116},
       number = {10},
          eid = {105002},
        pages = {105002},
          doi = {10.1103/PhysRevLett.116.105002},
archivePrefix = {arXiv},
       eprint = {1509.06601},
 primaryClass = {physics.flu-dyn},
       adsurl = {https://ui.adsabs.harvard.edu/abs/2016PhRvL.116j5002M}
}

@ARTICLE{Muller2000,
       author = {{M{\"u}ller}, Wolf-Christian and {Biskamp}, Dieter},
        title = "{Scaling Properties of Three-Dimensional Magnetohydrodynamic Turbulence}",
      journal = {\prl},
         year = 2000,
        month = jan,
       volume = {84},
       number = {3},
        pages = {475-478},
          doi = {10.1103/PhysRevLett.84.475},
archivePrefix = {arXiv},
       eprint = {physics/9906003},
 primaryClass = {physics.flu-dyn},
       adsurl = {https://ui.adsabs.harvard.edu/abs/2000PhRvL..84..475M}
}

@INPROCEEDINGS{Ng2007,
       author = {{Ng}, C.~S. and {Bhattacharjee}, A.},
        title = "{Anisotropic MHD turbulence}",
    booktitle = {Turbulence and Nonlinear Processes in Astrophysical Plasmas},
         year = 2007,
       editor = {{Shaikh}, Dastgeer and {Zank}, Gary P.},
       series = {American Institute of Physics Conference Series},
       volume = {932},
        month = aug,
    publisher = {AIP},
        pages = {137-143},
          doi = {10.1063/1.2778956},
archivePrefix = {arXiv},
       eprint = {1109.0974},
 primaryClass = {astro-ph.SR},
       adsurl = {https://ui.adsabs.harvard.edu/abs/2007AIPC..932..137N}
}

@ARTICLE{Perez2008,
       author = {{Perez}, Jean Carlos and {Boldyrev}, Stanislav},
        title = "{On Weak and Strong Magnetohydrodynamic Turbulence}",
      journal = {\apjl},
         year = 2008,
        month = jan,
       volume = {672},
       number = {1},
        pages = {L61},
          doi = {10.1086/526342},
archivePrefix = {arXiv},
       eprint = {0712.2086},
 primaryClass = {astro-ph},
       adsurl = {https://ui.adsabs.harvard.edu/abs/2008ApJ...672L..61P}
}

@ARTICLE{Politano1995,
       author = {{Politano}, H. and {Pouquet}, A.},
        title = "{Model of intermittency in magnetohydrodynamic turbulence}",
      journal = {\pre},
         year = 1995,
        month = jul,
       volume = {52},
       number = {1},
        pages = {636-641},
          doi = {10.1103/PhysRevE.52.636},
       adsurl = {https://ui.adsabs.harvard.edu/abs/1995PhRvE..52..636P}
}

@ARTICLE{Ripperda2021,
       author = {{Ripperda}, B. and {Mahlmann}, J.~F. and {Chernoglazov}, A. and {TenBarge}, J.~M. and {Most}, E.~R. and {Juno}, J. and {Yuan}, Y. and {Philippov}, A.~A. and {Bhattacharjee}, A.},
        title = "{Weak Alfv{\'e}nic turbulence in relativistic plasmas. Part 2. current sheets and dissipation}",
      journal = {Journal of Plasma Physics},
         year = 2021,
        month = oct,
       volume = {87},
       number = {5},
          eid = {905870512},
        pages = {905870512},
          doi = {10.1017/S0022377821000957},
archivePrefix = {arXiv},
       eprint = {2105.01145},
 primaryClass = {astro-ph.HE},
       adsurl = {https://ui.adsabs.harvard.edu/abs/2021JPlPh..87e9012R}
}

@ARTICLE{Saur2002,
       author = {{Saur}, J. and {Politano}, H. and {Pouquet}, A. and {Matthaeus}, W.~H.},
        title = "{Evidence for weak MHD turbulence in the middle magnetosphere of Jupiter}",
      journal = {\aap},
         year = 2002,
        month = may,
       volume = {386},
        pages = {699-708},
          doi = {10.1051/0004-6361:20020305},
       adsurl = {https://ui.adsabs.harvard.edu/abs/2002A&A...386..699S}
}

@ARTICLE{Schekochihin2022,
       author = {{Schekochihin}, Alexander A.},
        title = "{MHD turbulence: a biased review}",
      journal = {Journal of Plasma Physics},
         year = 2022,
        month = oct,
       volume = {88},
       number = {5},
          eid = {155880501},
        pages = {155880501},
          doi = {10.1017/S0022377822000721},
archivePrefix = {arXiv},
       eprint = {2010.00699},
 primaryClass = {physics.plasm-ph},
       adsurl = {https://ui.adsabs.harvard.edu/abs/2022JPlPh..88e1501S}
}

@ARTICLE{Shoda2019,
       author = {{Shoda}, Munehito and {Suzuki}, Takeru Ken and {Asgari-Targhi}, Mahboubeh and {Yokoyama}, Takaaki},
        title = "{Three-dimensional Simulation of the Fast Solar Wind Driven by Compressible Magnetohydrodynamic Turbulence}",
      journal = {\apjl},
         year = 2019,
        month = jul,
       volume = {880},
       number = {1},
          eid = {L2},
        pages = {L2},
          doi = {10.3847/2041-8213/ab2b45},
archivePrefix = {arXiv},
       eprint = {1905.11685},
 primaryClass = {astro-ph.SR},
       adsurl = {https://ui.adsabs.harvard.edu/abs/2019ApJ...880L...2S}
}

@ARTICLE{She1994,
       author = {{She}, Zhen-Su and {Leveque}, Emmanuel},
        title = "{Universal scaling laws in fully developed turbulence}",
      journal = {\prl},
         year = 1994,
        month = jan,
       volume = {72},
       number = {3},
        pages = {336-339},
          doi = {10.1103/PhysRevLett.72.336},
       adsurl = {https://ui.adsabs.harvard.edu/abs/1994PhRvL..72..336S}
}

@ARTICLE{Takamoto2016,
       author = {{Takamoto}, Makoto and {Lazarian}, Alexandre},
        title = "{Compressible Relativistic Magnetohydrodynamic Turbulence in Magnetically Dominated Plasmas and Implications for a Strong-coupling Regime}",
      journal = {\apjl},
         year = 2016,
        month = nov,
       volume = {831},
       number = {2},
          eid = {L11},
        pages = {L11},
          doi = {10.3847/2041-8205/831/2/L11},
archivePrefix = {arXiv},
       eprint = {1610.01373},
 primaryClass = {astro-ph.HE},
       adsurl = {https://ui.adsabs.harvard.edu/abs/2016ApJ...831L..11T}
}

@ARTICLE{Takamoto2017,
       author = {{Takamoto}, M. and {Lazarian}, A.},
        title = "{Strong coupling of Alfv{\'e}n and fast modes in compressible relativistic magnetohydrodynamic turbulence in magnetically dominated plasmas}",
      journal = {\mnras},
         year = 2017,
        month = dec,
       volume = {472},
       number = {4},
        pages = {4542-4550},
          doi = {10.1093/mnras/stx2292},
archivePrefix = {arXiv},
       eprint = {1709.00785},
 primaryClass = {astro-ph.HE},
       adsurl = {https://ui.adsabs.harvard.edu/abs/2017MNRAS.472.4542T}
}

@ARTICLE{Thompson1994,
       author = {{Thompson}, C.},
        title = "{A model of gamma-ray bursts.}",
      journal = {\mnras},
         year = 1994,
        month = oct,
       volume = {270},
        pages = {480-498},
          doi = {10.1093/mnras/270.3.480},
       adsurl = {https://ui.adsabs.harvard.edu/abs/1994MNRAS.270..480T}
}

@ARTICLE{Thompson1998,
       author = {{Thompson}, Christopher and {Blaes}, Omer},
        title = "{Magnetohydrodynamics in the extreme relativistic limit}",
      journal = {\prd},
         year = 1998,
        month = mar,
       volume = {57},
       number = {6},
        pages = {3219-3234},
          doi = {10.1103/PhysRevD.57.3219},
       adsurl = {https://ui.adsabs.harvard.edu/abs/1998PhRvD..57.3219T}
}

@ARTICLE{Toth2000,
       author = {{T{\'o}th}, G{\'a}bor},
        title = "{The {\ensuremath{\nabla}}{\textperiodcentered} B=0 Constraint in Shock-Capturing Magnetohydrodynamics Codes}",
      journal = {Journal of Computational Physics},
         year = 2000,
        month = jul,
       volume = {161},
       number = {2},
        pages = {605-652},
          doi = {10.1006/jcph.2000.6519},
       adsurl = {https://ui.adsabs.harvard.edu/abs/2000JCoPh.161..605T}
}

@ARTICLE{Verdini2018,
       author = {{Verdini}, Andrea and {Grappin}, Roland and {Alexandrova}, Olga and {Lion}, Sonny},
        title = "{3D Anisotropy of Solar Wind Turbulence, Tubes, or Ribbons?}",
      journal = {\apj},
         year = 2018,
        month = jan,
       volume = {853},
       number = {1},
          eid = {85},
        pages = {85},
          doi = {10.3847/1538-4357/aaa433},
archivePrefix = {arXiv},
       eprint = {1802.09837},
 primaryClass = {astro-ph.SR},
       adsurl = {https://ui.adsabs.harvard.edu/abs/2018ApJ...853...85V}
}

@ARTICLE{Wang2024,
       author = {{Wang}, Ru-Yue and {Zhang}, Jian-Fu and {Lu}, Fang and {Xiang}, Fu-Yuan},
        title = "{Exploring the intermittency of magnetohydrodynamic turbulence by synchrotron polarization radiation}",
      journal = {\aap},
         year = 2024,
        month = nov,
       volume = {691},
          eid = {A26},
        pages = {A26},
          doi = {10.1051/0004-6361/202450414},
archivePrefix = {arXiv},
       eprint = {2409.05739},
 primaryClass = {astro-ph.HE},
       adsurl = {https://ui.adsabs.harvard.edu/abs/2024A&A...691A..26W}
}

@ARTICLE{Xiao2025,
       author = {{Xiao}, Ya-Wen and {Zhang}, Jian-Fu and {Xu}, Siyao},
        title = "{Studying the diffusion mechanism of cosmic-ray particles}",
      journal = {\aap},
         year = 2025,
        month = jul,
       volume = {699},
          eid = {A317},
        pages = {A317},
          doi = {10.1051/0004-6361/202453340},
archivePrefix = {arXiv},
       eprint = {2506.15031},
 primaryClass = {astro-ph.HE},
       adsurl = {https://ui.adsabs.harvard.edu/abs/2025A&A...699A.317X}
}

@ARTICLE{Yan2004,
       author = {{Yan}, Huirong and {Lazarian}, A.},
        title = "{Cosmic-Ray Scattering and Streaming in Compressible Magnetohydrodynamic Turbulence}",
      journal = {\apj},
         year = 2004,
        month = oct,
       volume = {614},
       number = {2},
        pages = {757-769},
          doi = {10.1086/423733},
archivePrefix = {arXiv},
       eprint = {astro-ph/0408172},
 primaryClass = {astro-ph},
       adsurl = {https://ui.adsabs.harvard.edu/abs/2004ApJ...614..757Y}
}

@ARTICLE{Zhang2021,
       author = {{Zhang}, Jian-Fu and {Xiang}, Fu-Yuan},
        title = "{Energetic Particle Acceleration in Compressible Magnetohydrodynamic Turbulence}",
      journal = {\apj},
         year = 2021,
        month = dec,
       volume = {922},
       number = {2},
          eid = {209},
        pages = {209},
          doi = {10.3847/1538-4357/ac28ff},
archivePrefix = {arXiv},
       eprint = {2109.11357},
 primaryClass = {astro-ph.HE},
       adsurl = {https://ui.adsabs.harvard.edu/abs/2021ApJ...922..209Z}
}

@ARTICLE{Zhang2023,
       author = {{Zhang}, Jian-Fu and {Xu}, Siyao and {Lazarian}, Alex and {Kowal}, Grzegorz},
        title = "{Particle acceleration in self-driven turbulent reconnection}",
      journal = {Journal of High Energy Astrophysics},
         year = 2023,
        month = nov,
       volume = {40},
        pages = {1-10},
          doi = {10.1016/j.jheap.2023.08.001},
archivePrefix = {arXiv},
       eprint = {2308.07572},
 primaryClass = {physics.plasm-ph},
       adsurl = {https://ui.adsabs.harvard.edu/abs/2023JHEAp..40....1Z}
}

@ARTICLE{Zhao2024,
       author = {{Zhao}, Siqi and {Yan}, Huirong and {Liu}, Terry Z. and {Yuen}, Ka Ho and {Wang}, Huizi},
        title = "{Identification of the weak-to-strong transition in Alfv{\'e}nic turbulence from space plasma}",
      journal = {Nature Astronomy},
         year = 2024,
        month = jun,
       volume = {8},
        pages = {725-731},
          doi = {10.1038/s41550-024-02249-0},
archivePrefix = {arXiv},
       eprint = {2301.06709},
 primaryClass = {astro-ph.SR},
       adsurl = {https://ui.adsabs.harvard.edu/abs/2024NatAs...8..725Z}
}

@ARTICLE{Zrake2012,
       author = {{Zrake}, Jonathan and {MacFadyen}, Andrew I.},
        title = "{Numerical Simulations of Driven Relativistic Magnetohydrodynamic Turbulence}",
      journal = {\apj},
         year = 2012,
        month = jan,
       volume = {744},
       number = {1},
          eid = {32},
        pages = {32},
          doi = {10.1088/0004-637X/744/1/32},
archivePrefix = {arXiv},
       eprint = {1108.1991},
 primaryClass = {astro-ph.HE},
       adsurl = {https://ui.adsabs.harvard.edu/abs/2012ApJ...744...32Z}
}

@ARTICLE{Zrake2016,
       author = {{Zrake}, Jonathan and {East}, William E.},
        title = "{Freely Decaying Turbulence in Force-free Electrodynamics}",
      journal = {\apj},
         year = 2016,
        month = feb,
       volume = {817},
       number = {2},
          eid = {89},
        pages = {89},
          doi = {10.3847/0004-637X/817/2/89},
archivePrefix = {arXiv},
       eprint = {1509.00461},
 primaryClass = {astro-ph.HE},
       adsurl = {https://ui.adsabs.harvard.edu/abs/2016ApJ...817...89Z}
}
\bibliographystyle{aasjournal}

\end{document}